\documentclass[twocolumn,english,aps,prb,showpacs,superscriptaddress,amssymb,amsfonts]{revtex4}
\usepackage{color}
\usepackage{array}
\usepackage{amsmath}
\usepackage{amssymb}
\usepackage{epsfig}
\usepackage{graphicx}
\usepackage{wasysym}
\usepackage{bm}

\newcommand{\be}{\begin{equation}}
\newcommand{\ee}{\end{equation}}
\newcommand{\bea}{\begin{eqnarray}}
\newcommand{\eea}{\end{eqnarray}}

\begin{document}

\title{Quasi-particle Statistics and Braiding from Ground State Entanglement}

\author{Yi Zhang}
\affiliation{Department of Physics, University of California,
Berkeley, CA 94720, USA}

\author{Tarun Grover}
\affiliation{Department of Physics, University of California,
Berkeley, CA 94720, USA}

\author{Ari Turner}
\affiliation{University of Amsterdam, Science Park 904, P.O.Box 94485, 1090 GL Amsterdam, The Netherlands}

\author{Masaki Oshikawa}
\affiliation{Institute for Solid State Physics, University of Tokyo,
Kashiwa 277-8581, Japan}

\author{Ashvin Vishwanath}
\affiliation{Department of Physics, University of California,
Berkeley, CA 94720, USA}

\begin{abstract}

Topologically ordered phases are gapped states, defined by the
properties of excitations when taken around one another. Here we
demonstrate a method to extract the statistics and braiding of
excitations, given just the set of ground-state wave functions on a
torus. This is achieved by studying the Topological Entanglement
Entropy (TEE) on partitioning the torus into two cylinders. In this
setting, general considerations dictate that the TEE generally differs from that in trivial
partitions and depends on the chosen ground state. Central
to our scheme is the identification of ground states with minimum
entanglement entropy, which reflect the quasi-particle excitations
of the topological phase. The transformation of these states allows
for a determination of the modular $\mathcal{S}$ and $\mathcal{U}$
matrices which encode quasi-particle properties. We demonstrate our method by extracting the
modular $\mathcal{S}$ matrix of a SU(2) spin symmetric chiral spin
liquid phase using a Monte Carlo scheme to calculate TEE, and prove that the quasi-particles obey semionic
statistics. This method offers a route to a nearly complete
determination of the topological order in certain cases.

\end{abstract}

\maketitle

\section{Introduction}
Topologically ordered states are exotic gapped quantum phases of
matter that lie beyond the Landau symmetry breaking paradigm
\cite{wen2004}. Well known examples include fractional quantum Hall
states, gapped quantum spin liquids and quantum dimer models
\cite{wen2004, anderson1987, wen1990, Read89, wen1991, read1991,
senthil2000, sondhi}. These phases are not characterized by
correlations or local order parameters but rather by long range
entanglement in their ground-state wave functions \cite{kitaev2003}.
Emergent excitations in these phases can carry fractional quantum
numbers and, in two dimensions, realize non-trivial statistics on
exchanging identical particles or on taking one particle type around
another \cite{nayak}.  Two important hallmarks of topologically
ordered phases are ground-state degeneracy in a space with non-zero
genus and a finite topological entanglement entropy (TEE)
\cite{hamma2005, kitaev2006, levin2006}. In this paper we show that
combined together, these two properties can be used to extract key
aspects of a topological phase such as braiding rules and
topological spin of quasi-particles\cite{nayak}.

Taking a more general perspective, we note that a gapped phase of matter remains stable
as long as the excitation gap between the many body ground state(s) and the first excited state
remains non-zero. This implies that all `universal properties' associated with the gapped phase, which we define as the properties that remain invariant
with respect to the change of the underlying Hamiltonian while maintaining the gap, must be encoded
in the ground state wavefunction(s). The question of extracting such universal properties solely from the ground state wave-function(s)
becomes especially interesting for two-dimensional topologically ordered phases. In this case, the
universal properties are related to the anyonic character of \textit{gapped} excitations, such as braiding rules of
elementary quasi-particles. The preceding argument seems to employ that in a topological ordered phase, all robust properties of gapped excitations
are encoded in the ground state(s) itself. Such a point of view was taken by X.-G. Wen in Ref. \cite{wen1990} where the notion of non-abelian berry phase was introduced
to extract the braiding and statistics of quasi-particles. However, the idea in Ref. \cite{wen1990} requires one to have access to an \textit{infinite set of ground-states}
labeled by a continuous parameter, and is difficult to implement. 

Recently, it was found that the ground-state entanglement entropy of a two dimensional topologically ordered
phase in a disk-shaped region $A$ with a smooth boundary of length $L$ takes the
form $S_A = \alpha L - \gamma$, where the universal constant
$\gamma$ is the TEE \cite{hamma2005, kitaev2006, levin2006}. When, The constant $\gamma$ equals $\log(D)$ where $D  = \sqrt{\sum d^2_i} $ is the ``total
quantum dimension'' associated with the topological phase while $d_i$ is the quantum dimension of $i$'th quasiparticle type.
Unfortunately, the total quantum dimension $D$ only provides a partial characterization of topological order (for example, two distinct topological phases can have same
value of $D$). A natural question arises whether the quantum entanglement can be used to extract individual quantum dimensions $d_i$ and perhaps, the anyonic braiding
and statistics associated with the quasiparticles? As we show in this paper, the answer to this question is yes. We will only require the degenerate set of ground state wavefunctions on the torus. 

Recall that the ground state has topological degeneracy when the system is defined on a
topologically nontrivial manifold. It is generally assumed that TEE is a quantity solely determined by the
total quantum dimension $D$ of the underlying topological theory as
$\gamma=\log D$ independent of which ground state it is being calculated for.
However, this holds true \textit{only when the
boundary of the region $A$ consists of topologically trivial closed
loops}. If the boundary of region $A$ is non-contractible, for
example if one divides the torus into a pair of cylinders, \textit{generically the
entanglement entropy is different for different ground states}. Indeed
as shown in Ref.~\onlinecite{dong2008} for a class of topological
field theories, the TEE depends on the particular linear combination
of the ground states when the boundary of region $A$ contains
non-contractible loops.

In this paper, we first elaborate on the ground-state dependence of
entanglement entropy, focusing on the case of a partition of torus
into two cylinders. We present an argument based on the strong
subadditivity property of quantum information and show that the TEE
per connected boundary is not identical to that for a trivial
bipartition, such as a disc cut out of the torus, where TEE is $\log
D$. This is illustrated as an `uncertainty' relation, between
entropies for two different cylindrical bipartitions of the torus.

We also demonstrate this ground-state dependence numerically, and
calculate the entanglement entropy for the chiral spin liquid
\cite{kalmeyer} (CSL) wave function as different linear
superpositions of the two ground states (Figure \ref{fig4}) with the
Variational Monte Carlo (VMC) method \cite{gros1989, hastings2010,
frank2011}. The physical origin of the ground-state dependence of
TEE is made explicit by studying a $Z_2$ toric code model
\cite{kitaev2003}. We introduce the notion of \emph{minimum entropy
states (MESs)}, namely the ground states with minimal entanglement
entropy (or maximal TEE, since the TEE always
reduces the entropy) for a given bipartition. These states can be
identified with the quasi-particles of the topological phase and
generated by insertion of the quasi-particles into the cycle enclosed by
region A. For a generic lattice wave function with finite
correlation length, such as the CSL wave functions we study, a
nonlocal measurement like TEE is essentially to identify this basis
of MESs.

Having established the dependence of the TEE on the ground states, we
detail a procedure that uses this dependence to extract the key properties of quasi-particle
excitations by determining modular $\mathcal S$ and $\mathcal{U}$
matrices -- a vital characteristic of topological order
\cite{wen1990,wen93,kitaev_honey,nayak,yellowbook}. Unlike the TEE,
which failed to distinguish topological phases with the same total
quantum dimension, modular $\mathcal S$ and $\mathcal{U}$
matrices provide a finer characterization of topological order. In fact, they fully determine the quasi-particle self and mutual statistics
as well as the individual quantum dimensions of particles
in a topological phase. In particular, the element $\mathcal{S}_{ij}$ of the modular $\mathcal S$ matrix determines the mutual statistics of $i$'th
quasiparticle with respect to the $j$'th quasiparticle while the element $\mathcal{U}_{ii}$ of (diagonal) $\mathcal{U}$ matrix determines the
self-statistics (`topological spin') of the $i$'th quasiparticle. This procedure
requires just the set of ground states, although the results pertain
to the braiding and fusing of gapped excitations. The basic idea is to
relate MESs for different entanglement bipartitions of the torus. 
The MESs, which reflect
quasi-particle excitations, are determined using TEE. As an
application, we extract the modular $\mathcal{S}$ matrix of an
$SU(2)$ symmetric CSL, a lattice equivalence to a $\nu=1/2$ Laughlin
state, through TEE calculated with a recently developed VMC scheme.
For illustrative purposes, we discuss in the Appendix how our algorithm applies to Kitaev's
toric code model \cite{kitaev2003}, a zero correlation length phase,
and extract the modular $\mathcal{S}$ matrix and
$\mathcal{U}$ matrices.

At a practical level, the procedure outlined here suggests that entanglement entropy could
be used to numerically diagnose details of topological order beyond
the total quantum dimension
\cite{furukawa2007,haque2007,yao2010,isakov2011,frank2011}, which is a single
number susceptible to numerical error. An elegant different approach
to a more complete identification of topological order is through
the study of the entanglement spectrum \cite{HaldaneLi}. However we
note that requires the existence of edge states and may not be
applicable for topological phases like the Z$_2$ spin liquid.
Furthermore, it is possible to compute TEE using Monte Carlo
techniques on relatively larger systems \cite{isakov2011,frank2011}, as also done in this paper,
where the entanglement spectrum is not currently available.

\section{Ground State Dependence of Topological Entanglement
Entropy}\label{sec:GSdepend}

Given a normalized wave function $\left|\Phi\right\rangle$ and a
partition of the system into subsystems $A$ and $B$, one can trace
out the subsystem $B$ to obtain the reduced density matrix on
subsystem $A$:
$\rho_{A}=Tr_{B}\left|\Phi\right\rangle\left\langle\Phi\right|$. The
Renyi entropies are defined as:

\begin{eqnarray*}
S_{n}=\frac{1}{1-n}\log\left(Tr\rho^{n}_{A}\right)
\end{eqnarray*}

where $n$ is an index parameter. Taking the limit $n\rightarrow 1$,
$S_n$ recovers the definition of the usual von Neumann entropy. In
this paper we will often discuss the Renyi entropy with index $n=2$:
$S_{2}= - \log\left(Tr\left(\rho^{2}_{A}\right)\right)$ since it can
be calculated most easily with the VMC method\cite{frank2011} and at
the same time, captures all the information that we are interested
in.

For a gapped phase in 2D with topological order and a disc shaped
region $A$ with smooth boundary of length $L_{A}$, the Area Law of
the Renyi entropy gives:

\begin{eqnarray}
S_{n}=\alpha_n L_{A}-\gamma
\label{eq:topodefn}
\end{eqnarray}

where we have omitted the sub-leading terms. Although the
coefficient $\alpha_n$ of the leading `boundary law' term is
non-universal, the sub-leading constant $\gamma$, which is often
dubbed as the TEE, is universal and a robust property of the phase
of matter for which $|\Phi\rangle$ is the ground state. When region
$A$ has a disc geometry, it has been shown that $\gamma$ for
different degenerate ground states are identical and it is also
insensitive to the Renyi entropy index $n$ \cite{dong2008,
flammia2009}. It equals $\gamma=\log D$, where $D$ is the total
quantum dimension of the model \cite{kitaev2006, levin2006}, and
offers a partial characterization of the underlying topological
order.

However, when the subsystem $A$ takes a non-trivial topology, or
more precisely when the boundary of $A$ is non-contractible, TEE
contains more information \cite{dong2008}, as we will elaborate
further in this paper. For simplicity of illustration, throughout we
focus on the case when the two-dimensional space is a torus $T_2$
and the subsystem $A$ wraps around the $\hat{y}$ direction of the
torus and takes the geometry of a cylinder. For such a geometry, the
$n$'th Renyi entropy corresponding to the wave function
$|\Phi\rangle = \sum_j c_j |\Xi_j\rangle$ is given by: $S_n =
\alpha_n L_A - \gamma'_n$, where $|\Xi_j\rangle$ is a special basis
that we will describe in detail below and $\gamma'_n$
\cite{dong2008}:

\begin{eqnarray}
\gamma'_n \left(\{p_j\}\right) & = & 2 \gamma +
\frac{1}{n-1}\log\left(\underset{j}{\sum}p_{j}^{n}d_{j}^{2\left(1-n\right)}\right)
\label{tee_n}
\end{eqnarray}

Here $d_j\geq 1$ is the quantum dimension of the $j$th
quasi-particle and $p_j = |c_j|^2$. For Abelian anyons, $d_j=1$.
Note, $d_j$ shares the same subscript $j$ as the states
$|\Xi_j\rangle$ because the states $|\Xi_j\rangle$ can be obtained
by inserting a quasi-particle with quantum dimension $d_j$ (the
ground state degeneracy on the torus is equal to the number of
distinct quasi-particles). This equation shows that the TEE for this
geometry depends on the wave function through $\{p_j\}$ as well as
the Renyi index $n$, unlike the case with disc geometry.

What is the physical significance of the basis states
$|\Xi_j\rangle$? We claim that these are precisely the eigenstates
of the nonlocal operators defined on the entanglement cut, which
distinguish the topologically degenerate ground states. For example,
in the case of the quantum Hall \cite{dong2008} (Sec.
\ref{subsec:csltheory}), these states are the eigenstates of the
Wilson loop operator associated with the Chern-Simons gauge field
around the hole exposed by the entanglement cut. Similarly, for a
$Z_2$ gauge theory (Sec. \ref{subsec:z2} ), these are the states
with definite electric and magnetic field fluxes perpendicular to
the entanglement cut. For Abelian states, which have $d_j = 1$ for
all $j$ and are the focus of this paper, the entanglement entropy
associated with the states $|\Xi_j\rangle$ is minimum, i.e.
heuristically, the entanglement cut has the maximum `knowledge'
about these states. For this reason we name them {\em Minimum
Entropy States (MESs)}.

\subsection{Strong Subadditivity and Topological Entanglement Entropy on the Torus: An `uncertainty'
principle}

In this section we discuss the TEE for bipartitions of a torus into
two cylinders. This can be done by slicing the torus in two distinct
ways, along the vertical or horizontal directions. Intuitively, one
might expect both bipartitions would have the same TEE of $2\gamma$,
given the two disconnected boundaries of the cylinders. However,
very general considerations based on strong subadditivity of
von-Neumann entropy alone suggest that this expectation cannot be
correct. In practice, it is known that for a wide class of
topological phases, TEE of such nontrivial bipartitions indeed
depends on the ground state selected\cite{dong2008}. Here we do not
address ground-state dependence, rather we demonstrate that TEE
cannot be identical to its value for trivial bipartitions. It
invokes strong subadditivity, a deep property of quantum
information \cite{NielsenChuang}. This will allow us to come up with an uncertainty
principle, which constrains the amount of information we have when
we cut the torus in two orthogonal directions. Its advantage is that
it assumes almost nothing about the phase, except that it is gapped.

Consider the ground-state wave function of a gapped phase in two
dimensions and three non-overlapping subregions $A,\,B,\,C$. The
von-Neumann entropies $S$ follow the strong subadditivity
condition\cite{NielsenChuang}:

\begin{equation}
S_{ABC}+S_B-S_{AB}-S_{BC}\leq 0 \label{eq:strongsubadditivity}
\end{equation}

\begin{figure}
\begin{centering}
\includegraphics[scale=0.35]{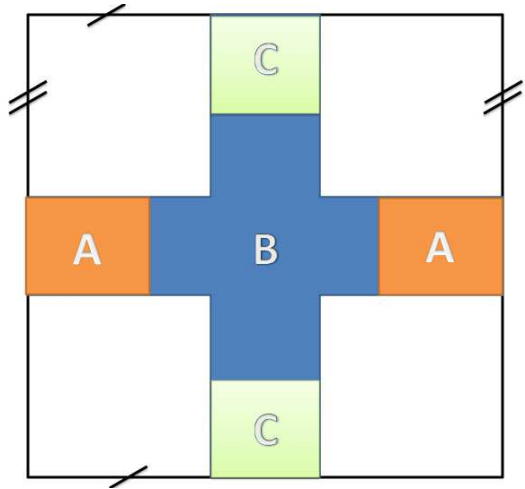}
\par\end{centering}
\caption{A torus (the top and bottom sides and left and right sides
are identified). Subregions $A,\,B,\,C$ are defined as shown.
Regions $A$ and $C$ are assumed to be well separated as compared to
the correlation length. The regions $AB$ and $BC$ correspond to
bipartitions of the torus into cylinders in orthogonal directions.}
\label{fig:torus}
\end{figure}

Note, this is only known to hold for von-Neumann entropies, not
Renyi entropies in general. Now, consider a torus with subregions
$A,\,B,\,C$ as shown in the Fig. \ref{fig:torus}. Let us decompose
the entropy into a part that arises from local contributions and a
non-local TEE $S=S^{\rm local}+S^{\rm topo}$. For a subregion with
the topology of a disc, the TEE is expected to be $S^{\rm
topo}=-\gamma$. Quite generally one can argue that $\gamma\geq0$
utilizing the strong subadditivity condition \cite{levin2006}. For
subregions defined on a simply connected surface, such as a disc,
the TEE is proportional to the number of connected components of the
boundary. If this was also true for the torus, we would expect
$S^{\rm topo}_{AB}=S^{\rm topo}_{BC}=-2\gamma$ (since they have a a
pair of boundaries). We now show this cannot be a consistent
assignment of TEE on the torus.

In order to isolate the topological part of the entropy, we assume
that the regions $A$ and $C$ are well separated compared to the
correlation length of the gapped ground state. Then, the local
contributions cancel in the combination above:  $S^{\rm
local}_{ABC}+S^{\rm local}_B-S^{\rm local}_{AB}-S^{\rm
local}_{BC}\rightarrow 0$. This can be argued following
Refs.\cite{levin2006,kitaev2006}. For example, consider a local
deformation near region $A$'s boundary far away from the other
regions. This change will be in $S^{\rm local}_{ABC}$, but a nearly
identical contribution will also appear in $S^{\rm local}_{AB}$,
since it only differs by the addition of a distant region. These
will cancel in the combination above. Thus, we can rewrite the Eqn.
\ref{eq:strongsubadditivity} as:

\begin{equation}
S^{\rm topo}_{ABC}+S^{\rm topo}_B-S^{\rm topo}_{AB}-S^{\rm
topo}_{BC}\leq 0 \label{eq:topo}
\end{equation}

This inequality implies the TEEs expected from the disc is not legal
for the torus.

For regions where the boundary is topologically trivial and
contractible (such as $ABC$ or $B$), one expects the TEE to be
independent of the surface on which they are defined, and hence
$S^{\rm topo}_{ABC}=S^{\rm topo}_B=-\gamma $. Only regions $AB$ and
$BC$, whose boundaries wrap around the torus, are sensitive to the
topology of the space they are defined on. Their TEEs satisfy:

\begin{equation}
\gamma_{BC}+\gamma_{AB}\leq 2\gamma
\end{equation}

where we have defined $S^{\rm topo}_{AB (BC)}=-\gamma_{AB (BC)}$.
Clearly this does not allow both the TEEs to to be $2\gamma$. In
fact if one of them attains its maximal disc value, the other must
vanish. Note, the TEE reduces the total entropy. Thus, when the
entropy of a cut along one of the cycles of the torus attains its
minimum value, i.e. we have most knowledge about the state on the
cut, then along the orthogonal direction, the entropy associated
with a cut must attain its maximal value, implying our knowledge is
the least. Therefore this can be thought of as an uncertainty
relation, between cuts that wrap around different directions of the
torus.

\subsection{Ground State Dependence of TEE in a Chiral Spin
Liquid}

In this subsection, to illustrate the state dependence of TEE, we
study the entanglement properties in a lattice model of an $SU(2)$
spin-symmetric CSL on a torus. The CSL has the same topological
order as the half filled Landau level $\nu = 1/2$ Laughlin state
\cite{kalmeyer, thomale} of bosons (these bosons can be thought of as
residing at the location of spin up moments), and has two-fold
degenerate ground states on the torus. The topological order in CSL can be confirmed by calculating its topological entanglement entropy (TEE) numerically using Monte Carlo
and verifying that it is non-zero and agrees with the  field theoretical predictions \cite{frank2011}. We note that here we are
working with a generic wave function in this phase defined on a
lattice, rather than with idealized zero correlation length states
or the topological field theory. This will introduce new conceptual
issues - in particular, the connection between MESs and lattice
ground states will be discussed.

We begin by reporting the results of a numerical experiment. We
extract TEE of linear combinations of the two ground states of the
CSL, and show that it indeed depends systematically on the chosen
linear combination, when the entanglement cut wraps around the
torus. We will then predict theoretically the dependence and find
excellent agreement as shown in Fig. \ref{fig4}.

\subsubsection{Numerical Study of Ground State Dependence of TEE in Gutzwiller Projected CSL states}
 \label{subsec:cslnumerics}

Wave functions of an SU(2) spin symmetric CSL are obtained in the
slave particle construction. We write the spins as bilinear in
fermions $\vec{S}=\frac12 f^\dagger_\sigma
[\vec{\sigma}]_{\sigma\sigma'}f_{\sigma'}$ and assume a chiral
d-wave state for the fermions. Operationally, the spin wavefunctions
are obtained by Gutzwiller projection of a $d_{x^2-y^2}+id_{xy}$
superconductor to one fermion per site. More technical details
regarding this wave function are in Appendix \ref{sec:varnWfn}. We
consider the system on a torus. Before projection, one can write
down different fermion states, by choosing periodic or anti-periodic
boundary conditions along $\hat x$ and $\hat y$ directions. These
boundary conditions are invisible to the spin degrees of freedom
which are bilinear in the fermions and lead to degenerate ground
states\cite{Read89}. We denote the ground states by the mean field
fluxes in $\hat x$ and $\hat y$ directions as
$\left|\varphi_{1},\varphi_{2}\right\rangle$, $\varphi_{1,2} = 0,
\pi$. The two fold degeneracy of the CSL implies that only two of
the four ground states $\left|0,0\right\rangle $, $\left|
\pi,0\right\rangle $, $\left|0,\pi\right\rangle $,
$\left|\pi,\pi\right\rangle $ are linearly independent. Here we
consider linear combinations of $\left|0,\pi\right\rangle $ and
$\left| \pi,0\right\rangle$, which we have numerically checked to be
indeed orthogonal for the system sizes that we consider:

\begin{equation}
\left|\Phi(\phi)\right\rangle =\cos\phi\left|0,\pi\right\rangle
+\sin\phi\left|\pi,0\right\rangle
\end{equation}

\begin{figure}
\begin{centering}
\includegraphics[scale=0.2]{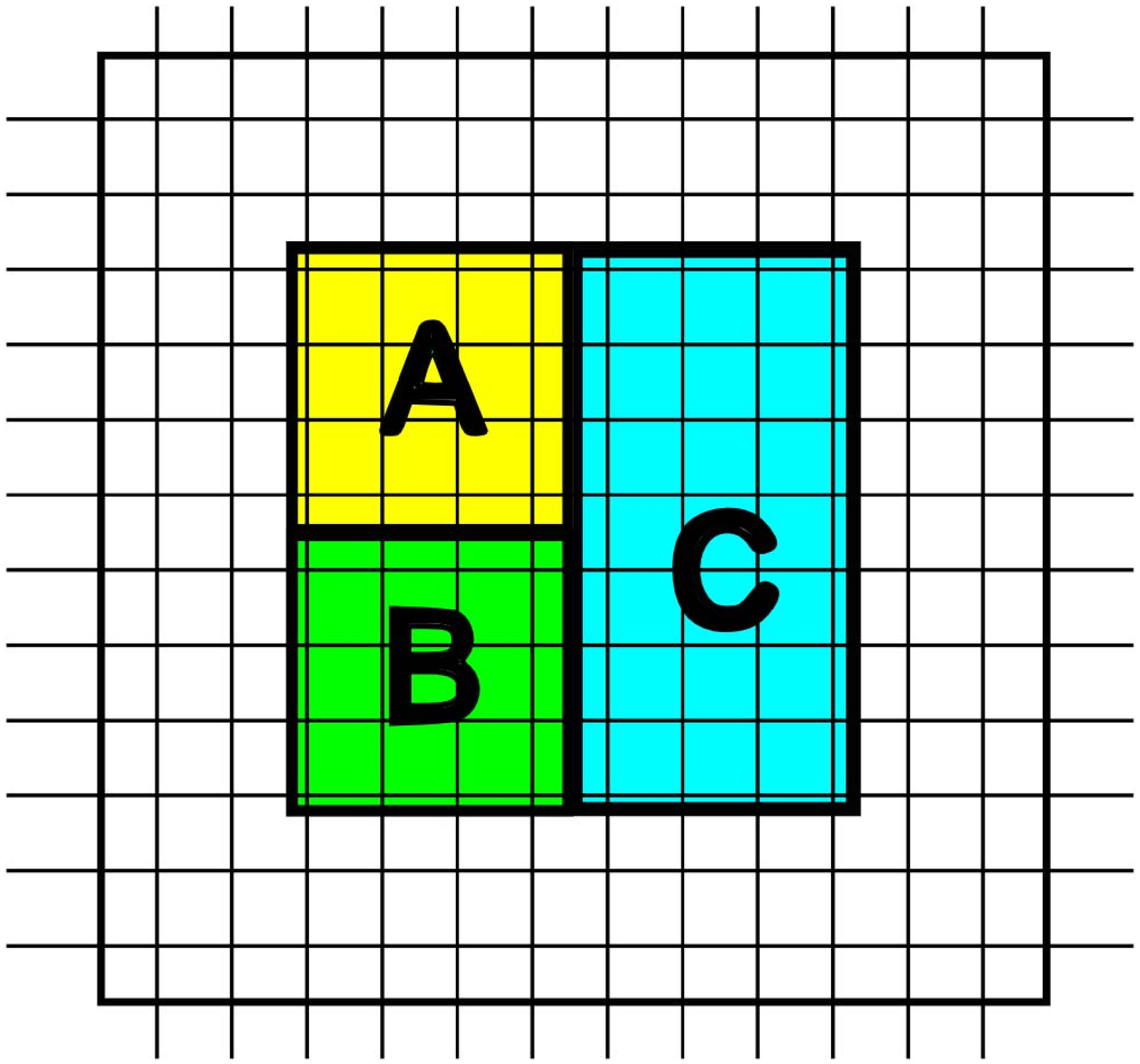}
\includegraphics[scale=0.2]{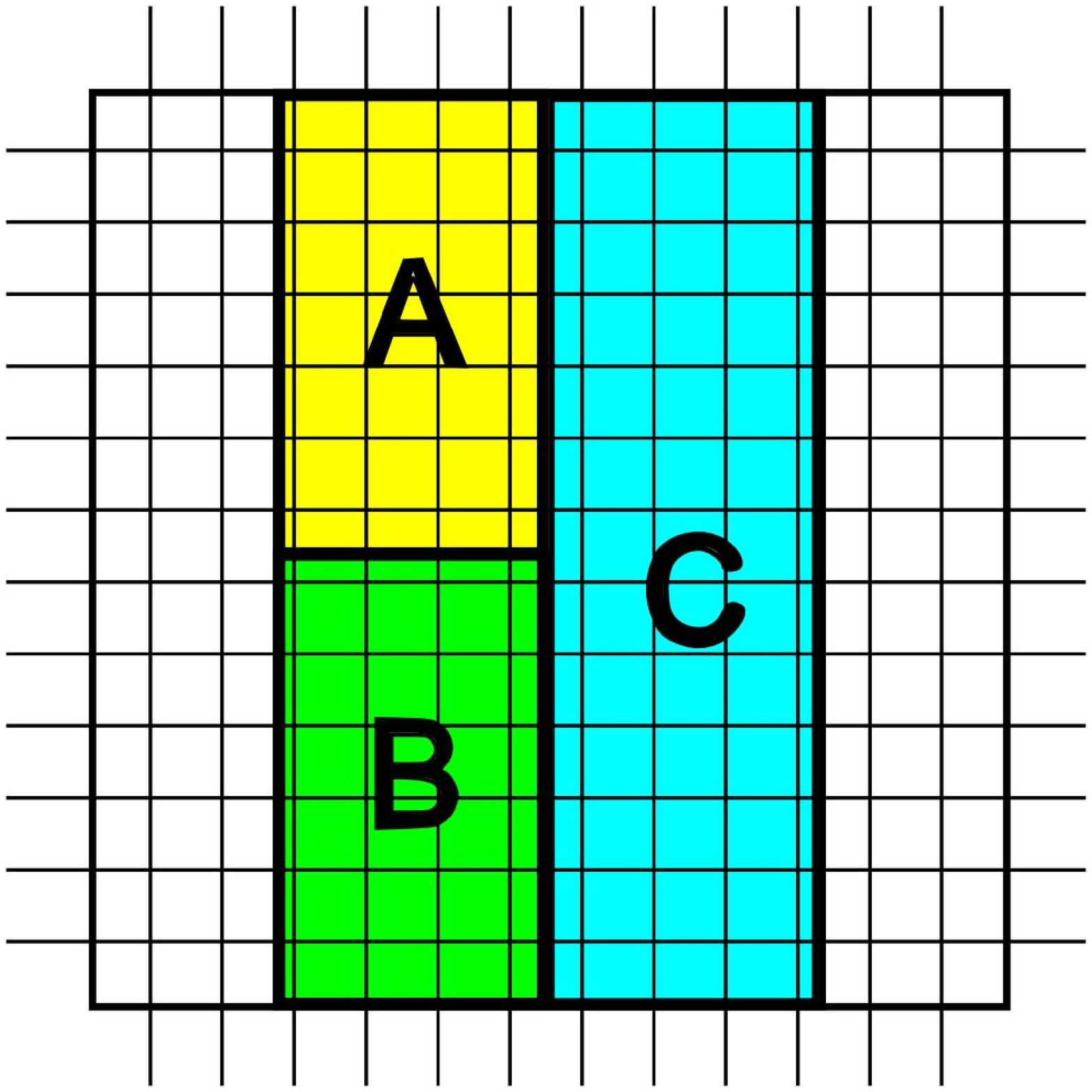}
\par\end{centering}
\caption{The separation of the system into subsystem $A$, $B$, $C$
and environment, periodic or antiperiodic boundary condition is
employed in both $\hat{x}$ and $\hat{y}$ directions. a: The
subsystem $ABC$ is an isolated square and the measured TEE has no
ground state dependence. b: The subsystem $ABC$ takes a non-trivial
cylindrical geometry and wraps around the $\hat{y}$ direction, and
TEE may possess ground stated dependence.} \label{fig3}
\end{figure}

We calculated TEE for the state $|\Phi\rangle$ using VMC method and
Gutzwiller projected wave functions based on Eqn.\ref{HamCSL}. An
efficient VMC algorithm which allows to study a linear combination
of Gutzwiller projected wave functions was developed and detailed in
Appendix \ref{sec:vmc}. To our knowledge, this is the first
numerical study to accomplish this.

The geometry and partition of the system are shown in Fig.
\ref{fig3}b. The total system size is 12 lattice spacings in both
directions with rectangles $A$ and $B$ being $6\times4$  and
rectangle $C$ $12\times4$. Note that the subsystems $AC$, $BC$,
$AB$, $C$ and $ABC$ all wrap around $\hat{y}$ direction so that
their TEE will all be equal (and denoted $\gamma'$). This is the
quantity we wish to access. For contractible subsystems $A$ and $B$
it remains the same as that expected for a region with a single
boundary, cut out of a topologically trivial surface (such as a
bigger disc) $\gamma$. We use the construction due to Kitaev and
Preskill \cite{kitaev2006} and effectively isolate the topological
contributions in the limit of small correlation length, by
evaluating the combination of entropies $S_{A}+S_{B}+S_{C} -
S_{AB}-S_{AC}-S_{BC}+S_{ABC}$. This combination is related to the
TEE by:

\begin{eqnarray}
-2\gamma+\gamma'&=&S_{A}+S_{B}+S_{C} \nonumber\\
&-&S_{AB}-S_{AC}-S_{BC}+S_{ABC} \nonumber\\
&=&2S_{A}-2S_{AC}+S_{ABC} \label{gammadifference}
\end{eqnarray}

\begin{figure}
\begin{centering}
\includegraphics[scale=0.35]{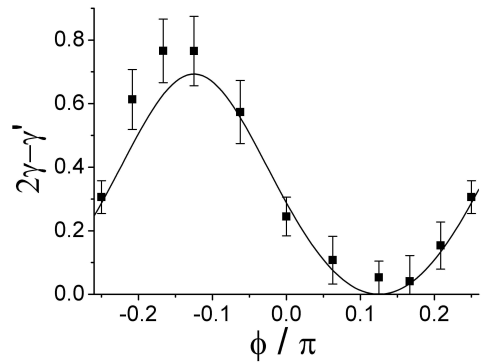}
\par\end{centering}
\caption{Numerically measured TEE $2\gamma - \gamma'$ for a CSL
ground state from linear combination $\left|\Phi\right\rangle
=\cos\phi\left|0,\pi\right\rangle +\sin\phi\left|\pi,0\right\rangle
$ as a function of $\phi$ with VMC simulations using geometry in
Fig. \ref{fig3}b. The solid curve is the theoretical value from Eqn.
\ref{eqntee7}. The periodicity is $\pi/2$.} \label{fig4}
\end{figure}

In the second line we have exploited symmetries of the construction
to reduce the problem to calculation of the Renyi entropy $S_{2}$ of
three regions $A$, $AC$ and $ABC$ for each $\phi$. To measure
$S_{2}$ numerically, we calculated the expectation value of a
$\rm{Swap_{A}}$ operator, see Ref\cite{frank2011} for an elaboration
of the method used. Our results for
$2\gamma-\gamma'\left(\phi\right)$ corresponding to different linear
combinations parameterized by $\phi$ are shown in Fig.\ref{fig4}.
This is one of the main results of this work.

We note the TEE strongly depends on the particular linear
combination chosen. The zero of the curve implies that the TEE
$\gamma'=2\gamma$, intuitive value for an entanglement cut with two
boundaries. The corresponding state is the MES. We note that the MES
occurs at a nontrivial angle. Understanding this requires connecting
the lattice states and the field theory which is done below. We
predict this angle to be $0.125\pi$ and the overall TEE dependence
to be Eqn. \ref{eqntee7}, which is plotted as the solid curve Fig.
\ref{fig4}, in rather good agreement with the numerical data.

\subsubsection{Theoretical Evaluation of Ground State Dependence of TEE in CSL wavefunctions}
 \label{subsec:csltheory}

A calculation of ground-state dependence of TEE involves two steps.
In the first step, we ask the following question: given a state expressed as a linear combination of MESs, what
is the expected TEE? For the CSL, this question has already been
answered by Ref.~\onlinecite{dong2008} that TEE for a state
$|\psi\rangle=a_1|1\rangle+a_2|2\rangle$ is

\begin{equation}
\gamma'-2\gamma=\log\left (  |a_1|^4+|a_2|^4 \right )
\end{equation}

where $|1\rangle,\,|2\rangle$ are MESs for cutting the torus in the direction in question.

Second, we need to understand the relation between the MES and the
physical states that appear in the Gutzwiller wave function. In
general it appears that the only way to identify MESs in a generic
wave function is by calculating the TEE. However, when the lattice
model has additional symmetry, that can also be used to identify MESs.
Here, we have a $12\times 12$ system defined on a square lattice and
we will exploit the $\pi/2$ rotation symmetry to establish a
connection between the flux states $|\varphi_1,\,\varphi_2\rangle$
of the Gutzwiller ansatz and the MESs.

The Gutzwiller projected ground states of the CSL,
$\left|0,0\right\rangle $ and $\left|\pi,\pi\right\rangle $ are
clearly invariant under a $\pi/2$ rotation symmetry \textit{upto a
phase factor}. A simple calculation shows that the
$\left|0,0\right\rangle $ state acquires phase factor $-1$ while the
$\left|\pi,\pi\right\rangle$ state acquires no phase under rotation.
Similarly, the states $\frac1{\sqrt{2}} \left
(\left|0,\pi\right\rangle\pm \left|\pi,0\right\rangle  \right )$
acquire a phase $\pm 1$ under rotation. Having established the
transformation of lattice states under rotation, we now study how
the MESs in the field theory respond to rotations. We will see that
$\pi/2$ rotation in the basis of the MESs is described by the
modular $\mathcal S$ matrix. The eigenvectors of the modular
$\mathcal S$ matrix will then be identified with lattice states that
are rotation eigenstates.

The CSL has the same topological order as the half filled Landau level $\nu = 1/2$
Laughlin state \cite{kalmeyer, thomale} of bosons. The field theory describing the topological order of a $\nu = 1/k$
Laughlin state is described by the following Chern-Simons action.  Note, here only the very
long wavelength degrees of freedom are retained:

\begin{eqnarray*}
S = \int
\frac{k}{4\pi}a_{\mu}\partial_{\upsilon}a_{\lambda}\epsilon^{\mu\upsilon\lambda}
\end{eqnarray*}

One can define the Wilson loop operators $T_1 = e^{i
\theta_{1}}=e^{i \oint a_{x}dx}$ and $T_2 = e^{i \theta_{2}}=e^{i
\oint a_{y}dy}$ around the two distinct cycles of the torus. In
terms of $\theta_i$, the action is given by

\begin{eqnarray*}
S= i\frac{k}{2\pi}\int dt\theta_{1}\dot{\theta_{2}}
\end{eqnarray*}

which implies that at the operator level
$\left[\theta_{1},\theta_{2}\right]= i\frac{2\pi}{k}$ or

\begin{eqnarray*}
T_1 T_2 = T_2 T_1 e^{2 \pi i /k}
\end{eqnarray*}

Owing to the above relation, there are $k$ orthogonal ground states
$| \psi_m \rangle $ that can be chosen to transform under $T_i$ as

\begin{eqnarray*}
T_2 |\psi_m \rangle & = & e^{2 \pi i (m-1)/k} |\psi_m\rangle \nonumber \\
T_1 |\psi_m \rangle & = & |\psi_{m+1}\rangle
\end{eqnarray*}

In the case of a CSL phase, $k=2$. Let us label the two degenerate
ground states as $\left(1,0\right)^T$ and $\left(0,1\right)^T$,
which are eigenstates of $T_2$:

\begin{eqnarray*}
T_2\left(1,0\right)^T&=&\left(1,0\right)^T \\
T_2\left(0,1\right)^T&=&-\left(0,1\right)^T \\
T_1\left(1,0\right)^T&=&\left(0,1\right)^T \\
T_1\left(0,1\right)^T&=&\left(1,0\right)^T \\
\end{eqnarray*}

The last two equations are due to the commutation relation $T_1 T_2
= -T_2 T_1$. It follows that the eigenstates of $T_1$ are
$\left(1,1\right)^T/\sqrt{2}$ and $\left(1,-1\right)^T/\sqrt{2}$.

\begin{figure}
\begin{centering}
\includegraphics[scale=0.5]{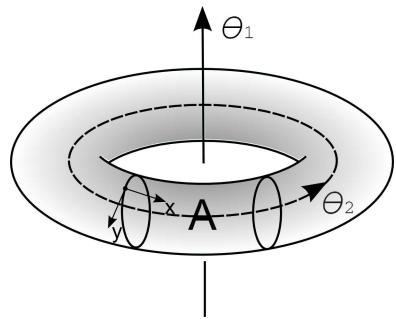}
\par\end{centering} \caption{The presence of two fluxes $\theta_{1}$ and $\theta_{2}$ on a torus. In cylindrical
cut  $A$ that wraps around $\hat{y}$ direction, only $\theta_{2}$ is
a measurable.} \label{fig1}
\end{figure}

The significance of the $T_{1,2}$ eigenstates is that they are
MESs\cite{dong2008}, for cuts whose boundaries are parallel to the
loops used to define $T_{1,2}$. This is because eigenstates of these
loop operators have a fixed value of flux enclosed within the
relevant cycle of the torus, which minimizes the entanglement
entropy for a parallel cut as shown in Fig. \ref{fig1}.

Now consider a $\pi/2$ rotation, under which $\theta_{1}\rightarrow
\theta_{2}$ and $\theta_{2}\rightarrow -\theta_{1}$ so $T_1
\rightarrow T_2$ and $T_2 \rightarrow T_1^{-1}=T_1$. Thus, the
matrix representing the effect of $\pi/2$ rotation for CSL in $T_2$
eigenstate basis is:

\begin{eqnarray} \mathcal{S}=\left(\begin{array}{cc}
\frac{1}{\sqrt{2}} & \frac{1}{\sqrt{2}}\\
\frac{1}{\sqrt{2}} & -\frac{1}{\sqrt{2}}\end{array}\right)
\label{smatCSL}
\end{eqnarray}

Note that we have used the symbol $\mathcal{S}$ for the above matrix
because it is indeed the modular $\mathcal S$ matrix of the
Chern-Simons topological quantum field theory corresponding to a
CSL. We recall that the modular $\mathcal S$ matrix transforms the
eigenstates of one Wilson loop operator $T_2$ to those of $T_1$. We
will return to the discussion of deriving $\mathcal{S}$ matrix for
CSL state using the entanglement properties of the ground states in
Sec. \ref{subsec:smatcsl}. Here we restrict ourselves to the
calculation of TEE for the CSL.

Since we are interested in the entanglement entropy with respect to
the cut shown in Fig. \ref{fig1}, let us represent all our states in
the basis of the eigenstates of $T_2$, i.e. the states $(0,1)$ and
$(1,0)$. Then, by matching eigenstates of the $\mathcal S$ matrix in
the above basis and rotation eigenstates of the lattice problem, we
conclude:

\begin{eqnarray*}
\left|\pi,0\right\rangle
&=&\left(\sin\frac{\pi}{8},\cos\frac{\pi}{8}\right)^T \nonumber \\
\left|0,\pi\right\rangle
&=&\left(\cos\frac{\pi}{8},-\sin\frac{\pi}{8}\right)^T
\end{eqnarray*}

We can now expand the general linear combination state $|\Phi(\phi)\rangle$ in MESs:

\begin{eqnarray}
\left|\Phi\right\rangle &=&\cos\phi\left|0,\pi\right\rangle
+\sin\phi\left|\pi,0\right\rangle \nonumber \\
&=&\left(\cos\left(\phi-\frac{\pi}{8}\right),\sin\left(\phi-\frac{\pi}{8}\right)\right)
\label{lin_comb} \label{phi}
\end{eqnarray}

Then, according to Eqn.~\eqref{tee_n}, theoretically one expects the
following expression for TEE:

\begin{eqnarray}
2\gamma-\gamma' &=&\log\frac{4}{3+\sin\left(4\phi\right)}
\label{eqntee7}
\end{eqnarray}

which is compared with the numerical data in Fig. \ref{fig4}. The
MES occur at the value of $\phi = \pi/8$ (mod $\pi/2$).

\subsection{Toric code and $Z_2$ spin liquid} \label{subsec:z2}

In this subsection we use the Kitaev's Toric code model
\cite{kitaev2003} as a pedagogical example to understand
ground-state dependence of TEE and the nature of the MESs for a $Z_2$
gauge theory.

\begin{figure}
\begin{centering}
\includegraphics[scale=0.2]{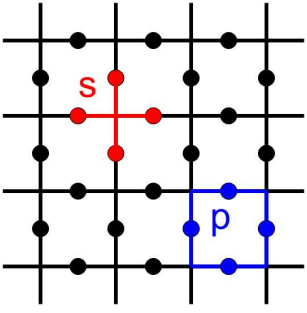}
\par\end{centering}
\caption{Illustration of a lattice of the toric code model, the
links spanned by star and plaquette are highlighted in red and blue,
respectively.} \label{fig5}
\end{figure}

Consider the toric code Hamiltonian of spins defined on the links of
a square lattice\cite{kitaev2003}:

\begin{eqnarray}
H = - \sum_{s} A_s - \sum_{p} B_p
\label{toriccode}
\end{eqnarray}

where $s$ and $p$ represent the links spanned by star and plaquette
as shown in the Fig.\ref{fig5}, and $A_s = \prod_{j \in s}
\sigma^{x}_j$, $B_p = \prod_{j \in p} \sigma^{z}_j$. Since all
individual terms in the Hamiltonian commute with each other, ground states are constructed from the simultaneous
eigenstates of all $A_s$ and $B_p$. Define the operator
$W^{z} (C)$ associated with a set of closed curves $C$ on the bonds of the lattice, as
follows

\begin{eqnarray}
W^{z}(C)&=&\prod_{j \in  C}\sigma_{j}^{z}
\end{eqnarray}

Then the ground state is an equal superposition of all possible loop
configurations:  $\sum_{C} W^{z}_{ab}(C) |\mbox{vac}_x\rangle$,
where   $|\mbox{vac}_x\rangle$ is a state with $\sigma_x=-1$ on
every site. The closed loops are interpreted as electric field lines
of the Z$_2$ gauge theory. We now consider two geometries, first the
cylinder and then the torus. The former case has a pair of
degenerate ground states, and is the simplest setting to demonstrate
state dependence of TEE.

\begin{figure}
\begin{center}
\includegraphics[width = 0.4 \textwidth]{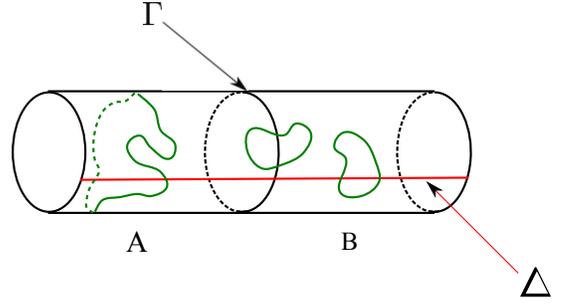}
\caption{ A snapshot of the ground state on the cylinder.
Closed-loop strings (``Z$_2$ electric fields'') can wrap around the
cylinder. The ground states are doubly degenerate, corresponding to
even and odd winding number sectors. The total number of string
crossings the cut $\Delta$ equals the winding number, modulo $2$.
The number of string crossings at the boundary $\Gamma$ is even in
the degenerate ground states. } \label{fig:cylinder}
\end{center}
\end{figure}

\subsubsection{Cylinder Geometry}
On a  cylinder, the Hamiltonian in Eqn.\ref{toriccode} leads to a
pair of degenerate ground states (the $A_s$ part of the Hamiltonian
is suitably modified at the boundary of the cylinder to only include
three links). The two normalized ground states  $|\xi_0 \rangle ,
|\xi_1 \rangle$,  are given by equal superpositions of electric
field loop configurations which have an even and odd winding number
around the cylinder respectively (see Fig. \ref{fig:cylinder}).
Consider now partitioning the cylinder into two cylindrical regions
$A$ and $B$. Then the Schmidt decomposition of these ground states
can be written as:

\begin{eqnarray}
\nonumber
 |\xi_0 \rangle &=&
\frac{1}{\sqrt{2 N_q}}
\sum_{\{ q_l\} }
\left(
| \Psi^A_{\{q_l\}, 0} \rangle
| \Psi^B_{\{q_l\}, 0} \rangle
+
| \Psi^A_{\{q_l\}, 1} \rangle
| \Psi^B_{\{q_l\}, 1} \rangle
\right)\\  \nonumber
 |\xi_1 \rangle &=&
\frac{1}{\sqrt{2 N_q}}
\sum_{\{ q_l\} }
\left(
| \Psi^A_{\{q_l\}, 0} \rangle
| \Psi^B_{\{q_l\}, 1} \rangle
+
| \Psi^A_{\{q_l\}, 1} \rangle
| \Psi^B_{\{q_l\}, 0} \rangle
\right)\\
\label{eq:decomp}
\end{eqnarray}

where the $N_q$ distinct configurations represented by $\{q_l\}$
denotes the electric field configurations at the cut. The number of
field lines crossing the cut is always even, since the ground state
is composed of closed loops. For trivial bipartitions, this exhausts
all terms in the Schmidt decomposition \cite{levin2006}. However,
given that the boundary of the cut is non-contractible, the
additional index $0,\,1$ appears which counts the parity of electric
field winding around the cylinder, within a partition. These are
correlated between the two partitions, for the fixed winding number
ground states. This is the key difference from a trivial
bipartition, leading to the ground-state dependence of TEE.

We now calculate the entanglement entropy associated with such a cut
for an arbitrary linear combination of these two ground states
$|\Psi\rangle=c_0|\xi_0 \rangle+c_1|\xi_1 \rangle$, with unit norm.
Using Eqn. \ref{eq:decomp} one can easily verify:

\begin{eqnarray}
\nonumber |\Psi \rangle =
\frac{1}{\sqrt{2 N_q}}
\sum_{\{ q_l\} }& &
 [(c_0+c_1)
| \Psi^A_{\{q_l\}, +} \rangle
| \Psi^B_{\{q_l\}, +} \rangle \\
+ &&
(c_0-c_1)| \Psi^A_{\{q_l\}, -} \rangle
| \Psi^B_{\{q_l\}, -} \rangle
]
\label{eq:decompCylinder}
\end{eqnarray}

where $| \Psi^{A(B)}_{\{q_l\}, \pm } \rangle = \left (|
\Psi^{A(B)}_{\{q_l\}, 0} \rangle\pm | \Psi^{A(B)}_{\{q_l\}, 1}
\rangle\right )/\sqrt{2}$. For a Schmidt decomposition $|\Psi\rangle
=\sum_a\sqrt{\lambda_a} |\Psi^A_a\rangle|\Psi^B_a\rangle$ the $n$th
Renyi entropy is given by: $S_n=\frac1{1-n}\log\left (  \sum_a
\lambda_a^n\right )$. We arrive at: $S_n=\frac1{1-n}\log
N_q^{1-n}[p_+^n+p_-^n]$, where $p_\pm=|c_0\pm c_1|^2/2$. Recognizing
that the closed loop constraint leads to $N_q=2^{L-1}$, where $L$ is
the length of the cut, and using the definition of TEE in Eqn.
\ref{eq:topodefn} we have:

\begin{equation}
\gamma'_n = \log 2 - \frac1{1-n}\log (p_+^n+p_-^n)
\end{equation}

Thus, for the electric field winding eigenstates $|\xi_{0,1}\rangle$
where $p_\pm=1/2$, the TEE vanishes. However, for their equal
superpositions when one of $p_+$ or $p_-$ vanishes, the TEE attains
its maximal value $\log 2$. These are eigenstates of the Wilson loop
operator that encircles the cylinder and measures the Z$_2$ magnetic
flux (vison number) threading it. An example of such a flux operator
is $F=\prod_{j \in Q}\sigma_{j}^{z}$, where $Q$ is a closed curve
that loops once around the cylinder, such as the boundary $\Gamma$
in Fig. \ref{fig:cylinder}. Since TEE reduces the entanglement
entropy, the maximum TEE states correspond to MESs. Why these MESs
are eigenstates of flux through the cylinder for this particular
cut? The number of electric field lines crossing the boundary
$\Gamma$ is always even. This constraint carries some information
and hence lowers the entropy by bringing in the standard TEE of
$\log{2}$. On the other hand, the topology of the cut boundary
$\Gamma$ allows for a determination of which magnetic flux sector
the cylinder is in. A state that is not an eigenstate of magnetic
flux through the cylinder leads to a loss of information and hence a
positive contribution to the total entanglement entropy (and reduces
TEE). This suggests that the MESs are eigenstates of loop operators
which can be defined parallel to the cut $\Gamma$. This is further
substantiated by the result for the torus case discussed below,
where they are simultaneous eigenstates of magnetic flux enclosed by
the cut and electric flux penetrating the cut.

\subsubsection{Torus Geometry}
The four degenerate ground states are distinguished by the even-odd
parity of the winding number of electric field lines around the two
cycles of the torus. The operator $W^{z}(C)$, which generates the
set of closed loops $C$ can be used to write the ground states:

\begin{eqnarray*}
|\xi_{ab}\rangle = \sum_{C} W^{z}_{ab}(C)
|\mbox{vac}_x\rangle
\end{eqnarray*}

where the subscript $a$ ($b$) takes on binary values $0,1$ and
denotes whether the loops $C$ belong to the even or odd winding
number sectors along the $\hat x$ ($\hat y$) direction, and
$|\mbox{vac}_x\rangle$ is a state with $\sigma_x=-1$ on every site.
The four ground states cannot be mixed by any local operator and
hence realize a $Z_2$ topological order. Let us consider a ground
state as the following linear combination:

\begin{eqnarray}
\left|\Psi\right \rangle = \sum_{a,b = 0,1} c_{a,b} \left|\xi_{ab}
\right\rangle \label{eqntee5}
\end{eqnarray}

We are interested in calculating entanglement entropy for the state
$\left|\Psi\right \rangle$ corresponding to the partition shown in
the Fig.\ref{fig1} and the dependence of TEE on parameters
$c_{a,b}$. After straightforward algebra (see details in Appendix
\ref{sec:mineez2}), one finds the following expression for subsystem
A with boundaries of length $L$:

\begin{eqnarray*} S_n = L \textrm{log}(2) - \gamma'_{n}
\end{eqnarray*}

where

\begin{eqnarray} \gamma'_{n} = 2\, \textrm{log}\,\left(2\right) - \frac{1}{1-n} \log \sum_{j = 1}^4 p_j^n
\label{pbcsgamma'}
\end{eqnarray}

and

\begin{eqnarray}
p_1 & = & \frac{|c_{00} + c_{01}|^2}{2} \nonumber \\
p_2 & = & \frac{|c_{00} - c_{01}|^2}{2} \nonumber \\
p_3 & = & \frac{|c_{10} + c_{11}|^2}{2} \nonumber \\
p_4 & = & \frac{|c_{10} - c_{11}|^2}{2} \label{prob_z2}
\end{eqnarray}

This is indeed consistent with Eqn.\ref{tee_n}, given that
$\gamma=\log D=\log 2$ and $d_j = 1$ for an Abelian topological
order with $D^2=4$ degenerate ground states. Further,
Eqn.\ref{pbcsgamma'} readily leads to the following four MESs:

\begin{eqnarray}
|\Xi_1 \rangle = \frac{1}{\sqrt{2}}(|\xi_{00}\rangle + |\xi_{01}\rangle) \nonumber \\
|\Xi_2 \rangle = \frac{1}{\sqrt{2}}(|\xi_{00}\rangle - |\xi_{01}\rangle) \nonumber \\
|\Xi_3 \rangle = \frac{1}{\sqrt{2}}(|\xi_{10}\rangle + |\xi_{11}\rangle) \nonumber \\
|\Xi_4 \rangle = \frac{1}{\sqrt{2}}(|\xi_{10}\rangle -
|\xi_{11}\rangle) \label{toricmes}
\end{eqnarray}

What is the physical significance of these four states being the
MESs? Similar to the cylinder geometry case, these states are the
simultaneous eigenstates of Wilson loop operators that encircles the
torus and measures the $Z_2$ magnetic and electric fluxes threading
it, as shown in Table \ref{table1} and Fig. We leave more detailed
algebra to Appendix \ref{sec:mineez2}.

When $\gamma'_{n}$ is maximized, the corresponding $S_n$ is
minimized, providing the maximum possible information about a given
state. Since the cut is made along $\hat y$, it can measure the
$Z_2$ magnetic and electric fluxes directed parallel to $\hat x$.
Hence the MES $|\Xi_\alpha\rangle$ with definite magnetic and
electric flux sectors, maximizes the TEE with $\gamma_{topo} = 2
\log(2)$, a contribution of $\log (2)$ from each of the two
boundaries. As one linear superposes different MESs
$|\Xi_\alpha\rangle$, the information obtained from measuring
magnetic and electric sectors becomes scrambled; especially, at the
extreme case of equal superposition of $|\Xi_\alpha\rangle$, all
information about the global quantum numbers has been lost and we
have $\gamma'=0$. This offers another example where MESs are the
eigenstates of loop operators defined on the cylinder from the
entanglement cut.

\begin{table}
\begin{tabular}{|c|c|c|c|}
\hline MES &$T_{y}$ & $F_{y}$ &quasi-particle\tabularnewline\hline
$\Xi_{1}$ & 0 & 0 & $1$\tabularnewline \hline $\Xi_{2}$ & 1 & 0 &
$m$\tabularnewline \hline $\Xi_{3}$ & 0 & 1 & $e$\tabularnewline
\hline $\Xi_{4}$ & 1 & 1 & $em$\tabularnewline \hline
\end{tabular}
\caption{List of $Z_2$ magnetic flux $T_y$, $Z_2$ electric flux
$F_y$ and corresponding quasi-particle of Wilson loop operator for
the four MESs $|\Xi_\alpha\rangle$ of the toric code with system
geometry in Fig.\ref{fig1}. The definitions of $T_y$ and $F_y$ are
in Appendix \ref{sec:mineez2}.} \label{table1}
\end{table}

{\bf Numerical Verification in a Generic Z$_2$ wavefunction:} To
provide an example of the state dependence of TEE in a $Z_2$
topologically ordered beyond the simple toric code model, we employ
our VMC method to calculate the TEE of an $SU(2)$ symmetric $Z_2$
spin liquid for a cylindrical bipartition of the torus. This is a
generic spin wave function, and has a finite correlation length
unlike the toric code. It  is obtained by Gutzwiller projecting a
mean-field BCS state on a square lattice (see Appendix
\ref{sec:varnWfn} for details).

\begin{figure}
\begin{centering}
\includegraphics[scale=0.3]{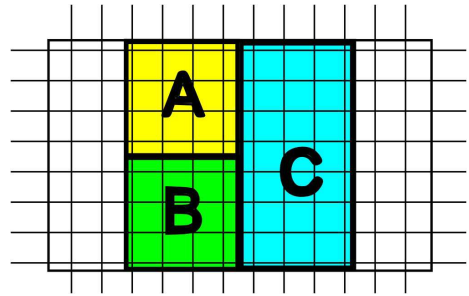}
\par\end{centering}
\caption{The separation of the system into subsystem $A$, $B$, $C$
and environment, periodic boundary condition is employed in both
$\hat{x}$ and $\hat{y}$ directions. The subsystem $ABC$ takes a
non-trivial cylindrical geometry and wraps around the $\hat{y}$
direction.} \label{fig6}
\end{figure}

For the $Z_2$ spin liquid wave function obtained by Gutzwiller
projecting the above mean-field ground state, a calculation of TEE
corresponding to a disc shaped region $A$ was performed in
Ref.\cite{frank2011}. It was established that the state is indeed
topologically ordered with the value of $\gamma \approx 0.584 \pm
0.089$, close to the expected theoretical value of $\gamma=\log 2
\approx 0.693$. Here we are concerned  with the calculation of TEE
for a cylindrical bipartition. We study systems with 12 lattice
spacings in the $\hat x$ direction and 8 lattice spacings in the
$\hat y$ direction. The subsystem separation scheme is similar to
Fig.\ref{fig3}b, where $A$ and $B$ are $4\times 4$ squares and $C$
is a $4\times 8$ rectangle, see Fig. \ref{fig6}. The TEE
$2\gamma-\gamma'$ can be evaluated in a similar way as
Eqn.\ref{gammadifference}.

By employing periodic boundary conditions in both directions, the
state we study has zero magnetic flux along both directions, but it
is not a MES. Instead, it is an equal superposition of two MESs
states that have zero magnetic flux and $0$ or $1$ electric flux
respectively through the cylinder. Therefore, according to
Eqn.\ref{pbcsgamma'}, with $p_1=p_3=1/2, p_2=p_4=0$ and Renyi index
$n=2$, one finds:

\begin{eqnarray*}
\gamma'&=& 2\log(2) + \log(\sum_{j = 1}^4 p_j^2)=\log\left(2\right)
\\ 2\gamma-\gamma'&=& \log\left(2\right)
\end{eqnarray*}

The VMC simulation yields $2\gamma-\gamma'=0.576\pm0.108$, close to
the expected theoretical value $\log(2) \sim 0.693$.
The smaller than expected values for
both $\gamma$\cite{frank2011, isakov2011} and $\gamma'$ are probably
due to quasi-particle excitations across a finite gap, causing
breaking of $Z_2$ electric field lines over the finite system size
we consider. Indeed, spin correlations decay slower for the $Z_2$
state, as compared to the CSL, which also arrives closer to its
expected value of TEE for both cylindrical bipartition as well as
for the disc shaped region $A$ \cite{frank2011}.

\section{Extracting Statistics from Topological Entanglement Entropy} \label{sec:topodata}

The modular $\mathcal{S}$ and $\mathcal{U}$ matrices describe the
action of certain modular transformations on the degenerate ground
states of the topological quantum field theory. On the other hand,
the braiding and statistics of quasi-particles are encoded in the
$\mathcal{S}$ and $\mathcal{U}$ matrices. For Abelian phases, the
$ij$'th entry of the $\mathcal{S}$ matrix corresponds to the phase
the $i$'th quasi-particle acquires when it encircles the $j$'th
quasi-particle. The $\mathcal{U}$ matrix is diagonal and the $ii$'th
entry corresponds to the phase the $i$'th quasi-particle acquires
when it is exchanged with an identical one. Since the MESs are the
eigenstates of the nonlocal operators defined on the entanglement
cut, the MESs are the canonical basis for defining $\mathcal S$ and
$\mathcal U$. The modular matrices are just certain unitary
transformations of the MES basis. As argued in Appendix
\ref{sec:mod}, the $\mathcal S$ matrix acts on MESs as an operator
that implements $\pi/2$ rotation while the $\mathcal US$ matrix
corresponds to $2 \pi/3$ rotation of MESs.

\subsection{Modular S-matrix of CSL from TEE} \label{subsec:smatcsl}

Let's consider the CSL wave functions studied in Sec.
\ref{subsec:cslnumerics}, and assume that we did not have any
information about the individual quantum dimensions or the modular
$\mathcal S$ matrix. The only information that is provided is the
two-fold degenerate ground-state wave functions $|\pi, 0 \rangle$
and $|0, \pi \rangle$. We construct the linear combination $|\Phi
\rangle$ as Eqn.\ref{lin_comb} and calculate its TEE for a
non-trivial bipartition as Fig.\ref{fig1} on a $\pi/2$ rotation
symmetric lattice. Consequently, we get the $2\gamma-\gamma'$
dependence on parameter $\phi$ in Fig.\ref{fig4}.

We notice that the minimum of the $2 \gamma' - \gamma$ attained is
approximately zero. According to Eqn.\ref{tee_n}, this implies that
at least one of the quantum dimensions $d_i$ should be $1$. Since
the total quantum dimension $D = \sqrt{d_0^2 + d_{1/2}^2} =
\sqrt{2}$, this implies that $d_0 = d_{1/2} = 1$. Also, we see that
the MES lies at $\phi \approx 0.14\pi$ by fitting
Fig.\ref{fig4} to Eqn.\ref{tee_n}.

For system with square geometry the $\mathcal{S}$ matrix describes
the action of $\pi/2$ rotation on the MESs. Since the two states
$|0,\pi\rangle$ and $|\pi,0\rangle$ transform into each other under
$\pi/2$ rotation, this implies that in the basis $\{|0,\pi\rangle,
|\pi,0\rangle\}$, the modular $\mathcal{S}$ matrix is given by the
Pauli matrix $\sigma_x$. To change the basis to MESs, we just need a
unitary transformation $V$ that rotates
$|0,\pi\rangle\,|\pi,0\rangle$ basis to the MESs basis. The $V$ is
determined by the fact that one needs to rotate
$|0,\pi\rangle\,|\pi,0\rangle$ basis by an angle $\approx 0.14 \pi$
to obtain MES (this is the numerically determined value, the exact
value being $\pi/8$). Therefore,

\begin{eqnarray*} \mathcal{S} = V^{\dagger} \left(\begin{array}{cc}
0 & 1 \\
1 & 0 \end{array}\right) V
\end{eqnarray*}

where

\begin{eqnarray*} V \approx \left(\begin{array}{cc}
\cos(0.14 \pi) & -\sin(0.14 \pi )e^{i\varphi} \\
\sin(0.14\pi) & \cos(0.14\pi )e^{i\varphi} \end{array} \right)
\end{eqnarray*}

from the two MESs: $(\cos(0.14 \pi), \sin(0.14 \pi))^T$ and
$(-\sin(0.14 \pi), \cos(0.14 \pi))^T$ and $\varphi$ is an undecided
phase. This yields the following value for the approximate modular
$\mathcal{S}$ matrix

\begin{eqnarray*} \mathcal{S} \approx \left(\begin{array}{cc}
\sin(0.28 \pi) & \cos(0.28 \pi)e^{i\varphi} \\
\cos(0.28 \pi)e^{-i\varphi} & -\sin(0.28 \pi) \end{array}\right)
\end{eqnarray*}

The existence of an identity particle requires positive real entries
in the first row and column and implies $\varphi=0$, which gives:

\begin{eqnarray*} \mathcal{S} \approx \left(\begin{array}{cc}
0.77 & 0.63 \\
0.63 & -0.77 \end{array}\right)
\end{eqnarray*}

Comparing this result with the exact expression in Eqn.
\ref{smatCSL}, we observe that even though the  $\mathcal{S}$ matrix
obtained using our method is approximate, some of the more important
statistics can be extracted rather exactly. The above $\mathcal{S}$
matrix tells us that the quasi-particle corresponding to $d_0 = 1$
does not acquire any phase when it goes around any other particle
and corresponds to an identity particle as expected, while the
quasi-particle corresponding to $d_{1/2} = 1$ has semion statistics
since it acquires a phase of $\pi$ when it encircles another
identical particle. Numerical improvements can further reduce the
error in pinpointing the MES and thereby leading to a more accurate
value of the $\mathcal S$ matrix. As another application, we study
the action of modular transformation on the MESs $|\Xi_\alpha
\rangle$ for the $Z_2$ gauge theory in Appendix \ref{sec:z2mods}.

\subsection{Algorithm for extracting modular $\mathcal{S}$ matrix from
TEE}\label{subsec:algo}

In the last subsection and Appendix \ref{sec:z2mods} we calculated
the modular $\mathcal S$ matrices for the CSL and $Z_2$ toric code
model respectively using the transformation properties of the basis states
$|\xi_{ab}\rangle $ under $\pi/2$ rotation transformations $\mathcal
R$ and translating it into the canonical basis of MESs
$|\Xi_i\rangle$: $\mathcal S = U^\dagger\mathcal R U$. However, such
$\pi/2$ rotation symmetry is not necessary. Even without symmetry,
it is possible to obtain the $\mathcal S$ matrix by studying the
modular transformation between certain nontrivial pair of sets of
MESs.

Starting from the definition, the modular $\mathcal{S}$ matrix has
the following expression:

\begin{equation}
\mathcal{S}_{\alpha\beta}=\frac{1}{D}\left\langle
\Xi_{\alpha}^{\hat{x}}|\Xi_{\beta}^{\hat{y}}\right\rangle
\label{eq:defineS}
\end{equation}

Here $D$ is the total quantum dimension and $\hat{x}$ and $\hat{y}$
are two directions on a torus. Eqn.\ref{eq:defineS} is just a
unitary transformation between the particle states along different
directions. In the case of a square system as in the last
subsection, the $\mathcal S$ matrix acts as a $\pi/2$ rotation on
the MES basis $\left|\Xi_{\beta}^{\hat{y}}\right\rangle$. In
general, however, $\hat{x}$ and $\hat{y}$ do not need to be
geometrically orthogonal, and the system does not need to be
rotationally symmetric, as long as the loops defining
$\left|\Xi_{\alpha}^{\hat{x}}\right\rangle$ and
$\left|\Xi_{\beta}^{\hat{y}}\right\rangle$ interwind with each
other. Therefore, the modular $\mathcal{S}$ matrix can be derived
even without any presumed symmetry of the given wave functions. Note
that there is an undetermined phase for each
$\left|\Xi_{\alpha}^{\hat{x}}\right\rangle$ and
$\left|\Xi_{\beta}^{\hat{y}}\right\rangle$, therefore a phase
freedom between the rows (columns), which may be fixed by the
existence of an identity particle.

Let's start with the two primitive vectors $w_{1}$ and $w_{2}$ and
determine the transformation of the MESs of $w_{2}$ to those of
$w_{2}'$ given by:

\begin{eqnarray}
w_{1}'=n_{1}w_{1}+m_{1}w_{2} \nonumber \\
w_{2}'=n_{2}w_{1}+m_{2}w_{2} \label{restrict1}
\end{eqnarray}

With $n_{1}m_{2}-m_{1}n_{2}=1$ by definition of the modular
transformation. We restrict $n_{2}=-1$, which means the cross
product:

\begin{equation}w_{2}\times w_{2}'=-w_{2}\times w_{1}=w_{1}\times w_{2}=A
\label{restrict2}
\end{equation}

$A$ is the surface area of the torus.

The corresponding modular matrix can be expanded as:

\begin{eqnarray*}\left(\begin{array}{cc}
n_{1} & 1-n_{1}m_{2}\\
-1 & m_{2}\end{array}\right)&=&\left(\begin{array}{cc}
1 & -n_{1}\\
 & 1\end{array}\right)\left(\begin{array}{cc}
 & 1\\
-1\end{array}\right)\left(\begin{array}{cc}
1 & -m_{2}\\
 & 1\end{array}\right)\\&=&U^{-n_{1}}SU^{-m_{2}}
 \end{eqnarray*}

Correspondingly, according to Appendix \ref{sec:mod} the
transformation:

\begin{equation*}\mathcal{R}=\mathcal{U}^{-n_{1}}\mathcal{S}\mathcal{U}^{-m_{2}}\end{equation*}

Because $\mathcal{U}$ matrix is diagonal by definition, its left
(right) matrix products only adds an additional phase factor to each
row (column) and can be eliminated. Therefore, without any argument
on the symmetry, the generalized algorithm:

\begin{enumerate}
\item Given a set of ground state wave functions
$\left|\xi_{\alpha}\right\rangle $, calculate the TEE of an
entanglement bipartition along $w_{2}$ direction, for a linear
combination $\left|\Phi\right\rangle =\sum
c_{\alpha}\left|\xi_{\alpha}\right\rangle $ . Search for the minimum
of TEE $2\gamma-\gamma'$ in the $c_{\alpha}$ parameter space. That
gives one MES $\left|\Xi_{\beta}\right\rangle $ and the
corresponding quantum dimension
$2\log\left(d_{\beta}\right)=2\gamma-\gamma'$. Note that the
existence of an identity particle ensures at least one minimum TEE
$2\gamma-\gamma'=0$.

\item iterate step 1 but with $c_{\alpha}$ in the parameter space
orthogonal to all previous obtained MESs
$\left|\Xi_{\beta}\right\rangle $. Continue this process until we
have the expressions for all $\left|\Xi_{\beta}\right\rangle $. This
gives a unitary transformation matrix $U_{1}$ with the
$\alpha\beta$'th entry being $c_{\alpha\beta}$, which changes the
basis from $\left|\xi_{\alpha}\right\rangle $ to
$\left|\Xi_{\beta}\right\rangle $. Note that there is a relative
$U(1)$ phase degree of freedom for each $|\Xi_\beta\rangle$.

\item Repeat step 1 and step 2 but with the entanglement cut along
$w_{2}'$ direction, which satisfies Eqn. \ref{restrict1} and Eqn.
\ref{restrict2}, and obtains the unitary transformation matrix
$U_{2}$.

\item The modular $\mathcal{S}$ matrix is given by $U_{2}^{-1}U_{1}$
except for an undetermined phase for each MES corresponding to a row
or a column. The existence of an identity particle that obtains
trivial phase encircling any quasi-particle helps to fix the
relative phase between different MESs, requiring the entries of the
first row and column to be real and positive. This completely
defines the modular $\mathcal{S}$ matrix.

\end{enumerate}

The above algorithm is able to extract the modular
transformation matrix $\mathcal{S}$ and hence braiding and mutual statistics of quasi-particle
excitations just using the ground-state wave
functions as an input. Further, there is no loss of generality for
non-Abelian phases, which can be dealt by enforcing the orthogonality condition in step
$2$ which guarantees that one obtains states with quantum dimensions $d_\alpha$ in an increasing
order.

In Appendix \ref{sec:z2mods} we take the square lattice toric code
model as an example once again, but without presuming any symmetry
of the system.

\subsection{Extracting other modular matrices from TEE}

In Appendix \ref{sec:z2mods}, we calculate the $\mathcal U$ matrix
for the $Z_2$ toric code model, given the simple action of $\mathcal
U$ on $|\xi_{ab} \rangle$. Though we were unable to find a general
algorithm for the $\mathcal U$ matrix, as we did for the the
$\mathcal S$ matrix in the last subsection, in the presence of certain
symmetries $\mathcal{U}$ can indeed be extracted given a set of ground-state wave functions
$|\xi_\alpha\rangle$. This is achieved by first calculating the action
$\mathcal R$ on the states $|\xi_\alpha\rangle$ under this symmetry operation, and
then translating it into the action on MESs. Specifically, the  corresponding modular matrix is given by $U^\dagger\mathcal R U$, where the unitary matrix $U$ is
obtained through the first two steps of the algorithm in the last
subsection.

The aforementioned symmetry to extract $\mathcal{S}$ matrix is the $\pi/2$ rotation, as shown in Sec. \ref{subsec:smatcsl} and the first
example in Appendix \ref{sec:z2mods}, but it
may be generalized to symmetries such as rotation of other angles
and even reflection symmetry (see Appendix \ref{sec:mod}). More interestingly, when
the symmetry operation $\mathcal R$ is a $2\pi/3$
rotation, one gets the $\mathcal US$ matrix. Hence, if one starts
with an arbitrary basis $| \xi_\alpha\rangle$ for the degenerate
ground state manifold of a topological order, the problem of
$\mathcal S$ and $\mathcal U$ matrices can be reduced to the
transformation property of chosen basis states $|\xi_\alpha\rangle$
under $\pi/2$ and $2 \pi/3$ rotations and the unitary transformation
that translates $|\xi_\alpha\rangle$ basis to the MESs
$|\Xi_\alpha\rangle$. To illustrate this point, we extract the $\mathcal{US}$
matrix for the $Z_2$ gauge theory in Appendix \ref{sec:z2mods} by putting the $Z_2$ toric code on triangular lattice which has $2 \pi/3$ rotation symmetry.

\section{Conclusion} \label{sec:concl}

In this paper, we studied two topologically ordered phases, the
chiral spin liquid (CSL) and the $Z_2$ spin liquid using VMC method
for Gutzwiller projected wave functions numerically and the $Z_2$
toric code model analytically, and showed that the topological
entanglement entropy (TEE) depends on the chosen ground state when
the entanglement bipartition is topologically nontrivial. We also
determined the minimum entropy states (MESs) and explained their
physical significance.

As an application of the physical significance of MESs, we suggested
an algorithm for extracting the modular transformation matrices that
determine the statistics such as quasi-particle self and mutual
statistics and individual quantum dimensions. These matrices also
determine the central charge of the edge state modulo 8
\cite{kitaev_honey}. Given the ground states, this algorithm
determines the topological order to a large extent. We note that Wen
proposed a different way to extract $\mathcal S$ and $\mathcal U$
matrices by calculating the non-abelian Berry phase \cite{wen1990}
which in practice may be difficult to implement, especially on a
lattice, since it requires calculating the degenerate ground states
$\psi_n$ of the system as function of the modular parameter $\tau =
\omega_2/\omega_1$ and calculating the derivatives such as  $\langle
\psi_n(\tau) |\partial_\tau |\psi_m(\tau)\rangle$.

We note that there may be cases where the $\pi/2$ and $2\pi/3$
rotations of the MESs may not be exactly identifiable with the
modular $\mathcal S$ and $\mathcal{US}$ matrices respectively. This
may happen, for example, when the particles have an internal angular
momentum which may cause the wavefunction to acquire an additional
phase upon rotation, over and above the phase due to underlying
topological structure. If the MESs correspond to spin-singlet
spin-liquid wave functions (such as CSL studied in this paper)
and/or string-net models (such as toric code model) where there is
no such internal structure, there should not be such additional
phase. Further, since all MESs are locally same, they should all
acquire same extra phase due to any local physics and therefore, the
extra phase may be separable from the topological phase.

For quantum Hall systems, because of the bulk-edge correspondence,
the fusion algebra and topological spin of the bulk quasi-particles
also determine the fusion rules and scaling dimensions for the
primary fields in the chiral CFT at the edge. Therefore, in the
context of quantum Hall systems, the entanglement entropy of the
ground state manifold determines robust features of the fields in
the corresponding edge CFT.

It would also be interesting to consider generalization of the
methods developed in our paper to higher dimensions. Discrete gauge
theories furnish the best known theories with long-range
entanglement in $D \ge 3$ dimensions and akin to $D=2$, they again
support degenerate ground states on the torus. In a recent paper
\cite{grov2011}, it was shown that these theories again have
non-zero TEE that is proportional to $\log(|G|)$, the number of
elements in the gauge group. A simple generalization of the method
developed in this paper shows that TEE for bipartition that has
non-contractible boundaries will again depend on the ground state,
and one will again find certain MESs that have the maximum knowledge
of the quantum numbers associated with an entanglement cut. Yet we
are not aware of simple generalization of modular transformations to
higher dimensions, the meaning of the matrix that relates MESs for
orthogonal entanglement cuts in higher dimensions requires further
investigation.

{\bf Acknowledgements:}
We thank Alexei Kitaev, Michael Levin, Chetan Nayak and Shinsei Ryu for helpful discussions.
We acknowledge support from NSF DMR 0645691.
A part of this research was performed at Kavli Institute for
Theoretical Physics, University of California Santa Barbara,
supported by the National Science Foundation
under Grant No. NSF PHY05-51164.
M. O. was supported in part by Grant-in-Aid for Scientific Research
(KAKENHI) No. 20102008.

\appendix

\section{Variational Monte Carlo method for a linear combination of wave-functions} \label{sec:vmc}

To calculate TEE for wave functions of different linear
combinations, it is important to establish a VMC algorithm for wave
function as $\left|\Phi\right\rangle
=\cos\phi\left|\Phi_{1}\right\rangle
+\sin\phi\left|\Phi_{2}\right\rangle $, where we assume
$\left|\Phi_{1}\right\rangle$ and $\left|\Phi_{2}\right\rangle$ are
properly normalized. In our case, $\left|\Phi_{1}\right\rangle $ and
$\left|\Phi_{2}\right\rangle $ are two degenerate ground states,
$\left\langle \alpha|\Phi_{1}\right\rangle $ and $\left\langle
\alpha|\Phi_{2}\right\rangle $ are single Slater determinants
products for each configuration $\left|\alpha\right\rangle$, making
$\left\langle \alpha|\Phi\right\rangle $ a sum of two Slater
determinants products. However, it may also be generalized to the
situation of any wave functions.

In the VMC scenario, the central quantity to evaluate in each Monte
Carlo step is the ratio of $\left\langle \alpha'|\Phi\right\rangle
/\left\langle \alpha|\Phi\right\rangle$, which now has the form:

\begin{eqnarray}
\frac{\left\langle \alpha'|\Phi\right\rangle }{\left\langle
\alpha|\Phi\right\rangle }= \frac{\cos\phi\left\langle
\alpha'|\Phi_{1}\right\rangle +\sin\phi\left\langle
\alpha'|\Phi_{2}\right\rangle }{\cos\phi\left\langle
\alpha|\Phi_{1}\right\rangle +\sin\phi\left\langle
\alpha|\Phi_{2}\right\rangle }\label{eqntee6}
\end{eqnarray}

It is usually much less costly to calculate ratio of $\left\langle
\alpha'|\Phi_1\right\rangle /\left\langle
\alpha|\Phi_1\right\rangle$ and $\left\langle
\alpha'|\Phi_2\right\rangle /\left\langle
\alpha|\Phi_2\right\rangle$ if $\left|\alpha\right\rangle$ and
$\left|\alpha'\right\rangle$ are locally different. For our case,
when $\left|\alpha\right\rangle$ and $\left|\alpha'\right\rangle$
differ only by one spin(electron) exchange, a much less costly and
more accurate algorithm may be implemented for the ratio of
determinants with only one different row or column. Unfortunately,
after linear superposing different $\left|\Phi_i\right\rangle$,
Eqn.\ref{eqntee6} no longer has such a privilege.

However, one can re express Eqn.\ref{eqntee6} as:

\begin{eqnarray*}
\frac{\left\langle \alpha'|\Phi\right\rangle }{\left\langle
\alpha|\Phi\right\rangle }= \frac{a+bc\cdot \tan\phi}{1+c\cdot
\tan\phi}
\end{eqnarray*}

where

\begin{eqnarray*}
a&=&\left\langle \alpha'|\Phi_{1}\right\rangle /\left\langle
\alpha|\Phi_{1}\right\rangle\\
b&=&\left\langle \alpha'|\Phi_{2}\right\rangle /\left\langle
\alpha|\Phi_{2}\right\rangle
\end{eqnarray*}

are again ratio of determinants and can be effectively evaluated,
and

\begin{eqnarray*}
c=\left\langle \alpha|\Phi_{2}\right\rangle /\left\langle
\alpha|\Phi_{1}\right\rangle
\end{eqnarray*}

can be efficiently kept track of with $c'=a^{-1}bc$ whenever the
update
$\left|\alpha\right\rangle\rightarrow\left|\alpha'\right\rangle$ is
accepted in a Monte Carlo step. In practice, numerical check should
be included to make sure error for $c$ does not accumulate too much
after a certain number of Monte Carlo steps.

This algorithm may be easily generalized to the linear combination
of $n$ wave functions, with the computational cost only $n$ times
that for a single wave functions.

\section{Variational wave functions for Chiral Spin Liquid and Z$_2$ Spin Liquid.}\label{sec:varnWfn}

\begin{figure}
\begin{centering}
\includegraphics[scale=0.3]{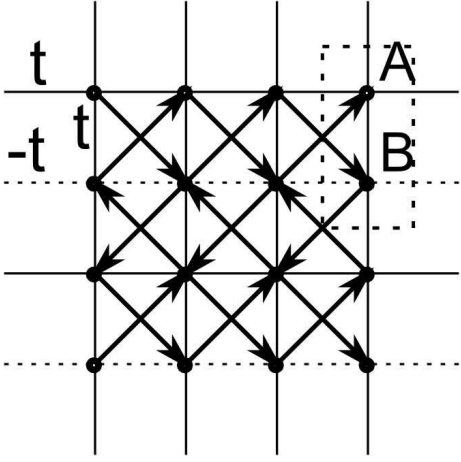}
\par\end{centering}
\caption{Illustration of a square lattice hopping model connected
with a $d+id$ superconductor. While the nearest neighbor hopping is
along the square edges with amplitude $t$ ($(-t)$ for hopping along
dash lines), the second nearest neighbor hopping is along the square
diagonal (arrows in bold), with amplitude $+i\Delta$ ($-i\Delta$)
when hopping direction is along (against) the arrow. The two
sublattices in the unit cell are marked as A and B.} \label{fig2}
\end{figure}

{\bf Chiral Spin Liquid From Gutzwiller Projection:} The lattice
wave function for the CSL states that we consider are obtained using
the slave-particle formalism by Gutzwiller projecting a $d+id$ BCS
state \cite{kalmeyer, thomale}. Specifically, we Gutzwiller project the
ground state of the following Hamiltonian of electrons hopping on a
square lattice at half filling:

\begin{eqnarray}
H=\underset{\left\langle ij\right\rangle
}{\sum}t_{ij}c_{i}^{\dagger}c_{j}+i\underset{\left\langle
\left\langle ik\right\rangle \right\rangle
}{\sum}\Delta_{ik}c_{i}^{\dagger}c_{k} \label{HamCSL}
\end{eqnarray}

Here $i$ and $j$ are nearest neighbors and the hopping amplitude
$t_{ij}$ is $t$ along the $\hat{y}$ direction and alternating
between $t$ and $-t$ in the $\hat{x}$ direction from row to row; and
$i$ and $k$ are second nearest neighbors connected by hoppings along
the square lattice diagonals, with amplitude $i\Delta_{ik}=i\Delta$
along the arrows and $i\Delta_{ik}=-i\Delta$ against the arrows, see
Fig.\ref{fig2}. The unit cell contains two sublattices A and B. This
model leads to a gapped state at half filling and the resulting
valence band has unit Chern number. This hopping model is equivalent
to a $d+id$ BCS state by an $SU(2)$ Gauge transformation. We take
$\Delta=0.5t$ to maximize the relative size of the gap and minimize
the finite size effect. Please refer to Ref\cite{frank2011} for
further details regarding the exact form of the wave function.

{\bf Z$_2$ Spin Liquid from the Gutzwiller Construction:}
\begin{eqnarray*}
H=-\underset{\left\langle ij\right\rangle
}{\sum}\left(\psi_{i}^{\dagger}\mu_{ij}\psi_{j}+h.c.\right)+\underset{i}{\sum}\psi_{i}^{\dagger}a_{0}^{l}\tau^{l}\psi_{i}
\end{eqnarray*}

where
$\psi_{i}=\left(f_{\uparrow},f_{\downarrow}^{\dagger}\right)^{T}$.
$\tau^{1,2,3}$ are Pauli matrices. The second term is related to
chemical potentials, we set $a_{0}^{2,3}=0$, with $a_{0}^{1}$ fixed
by the conditions $\left\langle
\psi^{\dagger}\tau^{1,2,3}\psi\right\rangle =0$. Matrices $\mu_{ij}$
connecting nearest and next nearest neighbors:

\begin{eqnarray*}
\mu_{i,i+x}&=&\mu_{i,i+y}=-\tau^{3}\\
\mu_{i,i+x+y}&=&\eta\tau^{1}+\lambda\tau^{2}\\
\mu_{i,i-x+y}&=&\eta\tau^{1}-\lambda\tau^{2}
\end{eqnarray*}

This mean field model is readily solvable, with dispersion:

\begin{eqnarray*}
E_{k} &=& \sqrt{\epsilon_{k}^{2}+\left|\Delta_{k}^{2}\right|} \\
\epsilon_{k}&=&2\left(\cos\left(k_{x}\right)+\cos\left(k_{y}\right)\right)\\
\Delta_{k}&=&2\eta\left[\cos\left(k_{x}+k_{y}\right)+\cos\left(k_{x}-k_{y}\right)\right]+a_{0}^{1}\\
&
&-2i\lambda\left[\cos\left(k_{x}+k_{y}\right)-\cos\left(k_{x}-k_{y}\right)\right]
\end{eqnarray*}

We choose $\eta=\lambda=1.5$ for a large gap and our calculation
estimates that the correlation length is as short as $\sim 1.3$
lattice spacings. Please refer to Ref\cite{frank2011} for more
details on the construction.

\section{Minimum entropy states of Toric code model on dividing torus} \label{sec:mineez2}

In this appendix we Schmidt-decompose the individual Toric code
ground states $\left|\Psi\right\rangle$ in Eqn.\ref{eqntee5} for the
bipartition of a torus in Fig.\ref{fig1}. It is helpful to introduce
a virtual cut $\Delta$ which wraps around the torus in the $\hat x$
direction, and define $\left|\Psi^{A\left(B\right)}_{{\{q_l\}},b}
\right\rangle$ as the normalized equal superposition of all the
possible configurations of closed-loop strings $C$ in the subsystem
$A$ ($B$) with the partition boundary condition specified by
${\{q_l=0,1\}}$, $l=1,2,...,L$(so L is the total length of the
boundary), and the number of crossings of the virtual cut $\Delta$
modulo 2 equals $b = 0,1$. The four ground states may now be
expanded as

\begin{eqnarray*} |\xi_{ab}\rangle &=& \frac{1}{\sqrt{2N_q}}\sum_{\{q_l\} \in \, a} \left(|\Psi^{A}_{\{q_l\},0} \rangle
|\Psi^{B}_{\{q_l\},b} \rangle \right. \\& &+
\left.|\Psi^{A}_{\{q_l\},1} \rangle |\Psi^{B}_{\{q_l\},(b+1)\,
\textrm{mod}\, 2} \rangle \right)
\end{eqnarray*}

Here $\{q_l\} \in \, a = 0$ $(1)$ denotes that the only even (odd)
number of crossings are allowed at the boundary $\Gamma_1$ (the
number of crossings at the other boundary $\Gamma_2$ must be same
modulo 2). $N_q=2^{L-2}$ equals the total number of valid boundary
conditions $\{q_l\} \in \, a$ in each parity sector.

We calculate entanglement entropy using the reduced density matrix.
Here, $\rho^A = \textrm{tr}_B |\Psi\rangle \langle \Psi| $ is
readily calculated,

\begin{eqnarray*}
\rho^{A}  & = & \frac{1}{2N_q} \sum_{\{q_l\} \in \, \textrm{even}}
\left[\left(|c_{00}|^2 + |c_{01}|^2\right)\left(|
\Psi^A_{\{q_l\},0}\rangle \langle \Psi^A_{\{q_l\},0}| \right.
\right. \\& & +
\left.|\Psi^A_{\{q_l\},1}\rangle \langle \Psi^A_{\{q_l\},1}| \right)  \nonumber \\
 &  &  \left. +  2 \,\textrm{Real}(c^{*}_{00} c_{01}) \left(| \Psi^A_{\{q_l\},0}\rangle \langle \Psi^A_{\{q_l\},1}| + |\Psi^A_{\{q_l\},1}\rangle \langle \Psi^A_{\{q_l\},0}|\right) \right]
 \nonumber \\
 &  & + \frac{1}{2N_q} \sum_{\{q_l\} \in \, \textrm{odd}} \left[\left(|c_{10}|^2 + |c_{11}|^2\right)\left(| \Psi^A_{\{q_l\},0}\rangle \langle \Psi^A_{\{q_l\},0}| \right.\right. \\& &+ \left.|
 \Psi^A_{\{q_l\},1}\rangle \langle \Psi^A_{\{q_l\},1}| \right)  \nonumber \\
 &  &  \left. +  2 \,\textrm{Real}(c^{*}_{10} c_{11}) \left(| \Psi^A_{\{q_l\},0}\rangle \langle \Psi^A_{\{q_l\},1}| + |\Psi^A_{\{q_l\},1}\rangle \langle \Psi^A_{\{q_l\},0}|\right) \right]
 \nonumber \\
& = & \frac{1}{2N_q} \sum_{\{q_l\} \in \, \textrm{even}} \left[
|c_{00} + c_{01}|^2  | \Psi^A_{\{q_l\},+}\rangle \langle
\Psi^A_{\{q_l\},+}| \right. \\& &+ \left.|c_{00} - c_{01}|^2 |
\Psi^A_{\{q_l\},-}\rangle \langle \Psi^A_{\{q_l\},-}| \right] \nonumber \\
&  & + \frac{1}{2N_q} \sum_{\{q_l\} \in \, \textrm{odd}} \left[
|c_{10} + c_{11}|^2 | \Psi^A_{\{q_l\},+}\rangle \langle
\Psi^A_{\{q_l\},+}|  \right. \\& &+ \left.|c_{10} - c_{11}|^2 |
\Psi^A_{\{q_l\},-}\rangle \langle \Psi^A_{\{q_l\},-}| \right] \nonumber \\
\end{eqnarray*}

Here $|\Psi^A_{\{q_l\}, \pm}\rangle = \frac{1}{\sqrt{2}} \left(
|\Psi_{\{q_l\},0}^ A \rangle \pm \Psi_{\{q_l\},1}^ A \rangle
\right)$ and hold the orthogonal condition.

From the above expression, it immediately follows that the Renyi
entanglement entropy $S_n$ is given by Eqn.\ref{pbcsgamma'}:

\begin{eqnarray*}
S_{n}&=&\frac{1}{1-n}\log\left(Tr\rho^{n}_{A}\right)\\
&=&\frac{1}{1-n}\log\left(\left(\frac{1}{2N_q}\right)^n\cdot
N_q\left(\sum_{j = 1}^4 (2p_j)^n\right)\right)\\
&=&\log N_q + \frac{1}{1-n}\log \sum_{j = 1}^4 p_j^n \\
&=&L \log 2 - \left(2\log 2 + \frac{1}{n-1}\log \sum_{j = 1}^4
p_j^n\right)
\end{eqnarray*}

where $p_j$ are defined in Eqn. \ref{prob_z2}.

To understand the nature of the corresponding MES in Eqn.
\ref{toricmes}, we first discuss the quasi-particle excitations of
the Toric code model. Imagine acting a string operator defined on
the links of the lattice

\begin{eqnarray*}
W^{z}(O)&=&\prod_{j \in  O}\sigma_{j}^{z}
\end{eqnarray*}

Now $W^{z}(O) |\mbox{vac}_x\rangle$ is an excited states and still
an eigenstate of $A_s$ and $B_p$, with $A_s=-1$ at the two ends of
$O$. We may regard them as electric charge quasi-particles that cost
a finite energy to create and the string connecting them as an
electric field line. To return to the ground state, the electric
charges need to be annihilated with each other. One way to do this
is to wrap the open string $O$ parallel to $\hat{x}$ around the
cycle of the torus. $O$ becomes a closed loop $C$, yet this changes
the parity of electric field winding number along $\hat{x}$. We
define the electric charge loop operator that insert an additional
electric field in the $\hat x$($\hat y$) direction by the above
procedure as a $Z_2$ electric flux insertion operator $T_x$($T_y$).

\begin{eqnarray}
T_x|\xi_{1b}\rangle &=& |\xi_{0b}\rangle \nonumber\\
T_x|\xi_{0b}\rangle &=& |\xi_{1b}\rangle \nonumber\\
T_y|\xi_{a1}\rangle &=& |\xi_{a0}\rangle \nonumber\\
T_y|\xi_{a0}\rangle &=& |\xi_{a1}\rangle \label{eq:t-act}
\end{eqnarray}

There is also a magnetic field, which determines the phase of the
electric charge as it moves. In particular, when there is a magnetic
field along the $\hat y$ direction of the torus of $1$($0$) total
flux($mod2$), the electric charge picks up a $-$($+$) sign traveling
around the loop around the $\hat x$ direction, and similarly for the
magnetic field along the $\hat x$ direction. Denoting the insertion
operator of such $Z_2$ magnetic flux as $F_y$ and $F_x$, the loop
operators of the magnetic charge (vison), we have,

\begin{eqnarray*}
T_x F_y = -F_y T_x\nonumber\\
T_y F_x = -F_x T_y
\end{eqnarray*}

They suggest that $T_x$($T_y$) is the magnetic flux measuring
operator in the $\hat y$($\hat x$) direction and $F_x$($F_y$) is the
electric flux measuring operator in the $\hat y$($\hat x$)
direction. Note that both electric and magnetic flux are defined
modulo 2 in correspondence with the $Z_2$ gauge theory. After simple
algebra,

\begin{eqnarray}
F_y|\xi_{ab}\rangle &=& (-1)^a|\xi_{ab}\rangle \nonumber\\
F_x|\xi_{ab}\rangle &=& (-1)^b|\xi_{ab}\rangle \label{eq:f-act}
\end{eqnarray}

Compare Eqn. \ref{eq:t-act} and \ref{eq:f-act} with Eqn.
\ref{toricmes}, we arrive at the conclusions listed in Table
\ref{table1}.

\section{Modular Transformations} \label{sec:mod}

The $\mathcal{S}$ and $\mathcal{U}$ matrices describe the action of
modular transformations on the degenerate ground states of the
topological quantum field theory on a torus. For Abelian phases, the
$ij$'th entry of the $\mathcal{S}$ matrix corresponds to the phase
the $i$'th quasi-particle acquires when it encircles the $j$'th
quasi-particle. The $\mathcal{U}$ matrix is diagonal and the $ii$'th
entry corresponds to the phase the $i$'th quasi-particle acquires
when it is exchanged with an identical one. Let us first review the
geometric meaning of these transformations. Labeling our system by
complex coordinates $z = x + i y$, the torus may be defined by the
periodicity of $\omega_1$ and $\omega_2$ along the two directions
$\hat{e_1}$ and $\hat{e_2}$ (need not to be orthogonal) i.e. $ z
\equiv z + \omega_1 \equiv z + \omega_2$. Now consider a
transformation

\begin{equation}
 \left(
\begin{array}{c}
\omega_1 \\
\omega_2 \\
\end{array} \right) \rightarrow
\left(
\begin{array}{c}
\omega'_1 \\
\omega'_2 \\
\end{array} \right) =
\left(
\begin{array}{cc}
a & b \\
c & d \\
\end{array} \right)
\left(
\begin{array}{c}
\omega_1 \\
\omega_2 \\
\end{array} \right)
\label{modular} \end{equation}

where $a,b,c,d \in \mathbb{Z}$. Since our system lives on a lattice,
the inverse of the above matrix should again have integer
components, hence the determinant $ad - bc = 1$. One can show that
matrices with these properties form a group, called $SL(2,
\mathbb{Z})$. Interestingly, all the elements in this group can be
obtained by a successive application of the following two generators
of $SL(2, \mathbb{Z})$:

\begin{itemize}
\item $S  = \left(
\begin{array}{cc}
0 & 1 \\
-1 & 0 \\
\end{array} \right)$. This transformation corresponds to $\omega_1 \rightarrow \omega_2$ and $\omega_2 \rightarrow -\omega_1$ and therefore, for a square
geometry corresponds to rotation of the system by $90^o$.

\item $U = \left(
\begin{array}{cc}
1 & 1 \\
0 & 1 \\
\end{array} \right)$. Under this transformation $\omega_1 \rightarrow \omega'_1 =
\omega_1 + \omega_2$ and $\omega_2 \rightarrow \omega'_2 =
\omega_2$. Consider a loop on the torus with winding numbers $n_1$
and $n_2$ along $\omega_1$ and $\omega_2$ directions. By definition
of the $U$ transformation, the winding numbers in the transformed
basis:

\begin{eqnarray*}
& & n_1 \omega_1 + n_2 \omega_2  \\
& = & n_1(\omega'_1 - \omega'_2) + n_2 \omega'_2 \\
& = & n'_1 \omega'_1 + n'_2 \omega'_2
\end{eqnarray*}

where $n'_1 = n_1$ and $n'_2 = n_2 - n_1$ are the winding numbers
along the $\omega'_1$ and $\omega'_2$ directions.

\end{itemize}

The transformation properties of the resulting MESs under modular
transformations would yield the desired $\mathcal S$ and $\mathcal
U$ matrices. Further, for a symmetry transformation of $F(S,U)$ on
$(\omega_1, \omega_2)^T$, the corresponding modular transformation
on MESs would yield the modular $\mathcal{F(S,U)}$ matrix.

In the main text, we have obtained $\mathcal S$ and $\mathcal U$
matrices for the toric code model from the action of these
transformations on the basis states $|\xi_{ab}\rangle$. We now show
that one can also obtain the $\mathcal US$ matrix by studying the
action of $2\pi/3$ rotation $R_{2\pi/3}$ on the MESs (provided that
$R_{2\pi/3}$ is symmetry of the model). To see this, consider a
triangular lattice that is defined by two lattice vectors (complex
numbers) $\omega_1, \omega_2$ with $\omega_1 = (1,0)$ and $\omega_2
= (1/2, \sqrt{3}/2)$. The transformation of our interest is the
transformation of $\omega_1, \omega_2$ under $R_{2\pi/3}$ rotation:
$\omega_1 \rightarrow \omega'_1 = - \omega_1 + \omega_2$ and
$\omega_2 \rightarrow \omega'_2 = -\omega_1$. Therefore, one can
write the $R_{2\pi/3}$-matrix

\be R_{2\pi/3} = \left(
\begin{array}{cc}
-1 & 1 \\
-1 & 0 \\
\end{array} \right)
\ee

This matrix belongs to the group $SL(2, \mathbb{Z})$ and simple
algebra shows that $R_{2\pi/3} = US$. One may also check that
$R_{2\pi/3}^3 = 1$ as one might expect. Therefore, knowing the
action of $R_{2\pi/3}$ on the MESs would lead to the $\mathcal{US}$
matrix.

\section{Modular matrices of $Z_2$ gauge theory by transforming
minimum entropy states}\label{sec:z2mods}

Let's study the action of modular transformation on the MESs
$|\Xi_\alpha \rangle$ for the $Z_2$ gauge theory in Sec.
\ref{subsec:z2} and compare the resulting modular matrices with the
known results.

First consider a $\pi/2$ rotation symmetric square sample. Under
$\pi/2$ rotation, $|\xi_{ab}\rangle \rightarrow |\xi_{ba}\rangle$.
According to Eqn.\ref{toricmes}, the transformation for the MESs
$|\Xi_\alpha \rangle$ for cuts along $\hat y$:

\begin{eqnarray*}
|\Xi_1 \rangle \rightarrow \frac{1}{2} \left( |\Xi_1 \rangle + |\Xi_2 \rangle + |\Xi_3 \rangle + |\Xi_4 \rangle  \right) \\
|\Xi_2 \rangle \rightarrow \frac{1}{2} \left( |\Xi_1 \rangle + |\Xi_2 \rangle - |\Xi_3 \rangle - |\Xi_4 \rangle  \right) \\
|\Xi_3 \rangle \rightarrow \frac{1}{2} \left( |\Xi_1 \rangle - |\Xi_2 \rangle + |\Xi_3 \rangle - |\Xi_4 \rangle  \right) \\
|\Xi_4 \rangle \rightarrow \frac{1}{2} \left( |\Xi_1 \rangle - |\Xi_2 \rangle - |\Xi_3 \rangle + |\Xi_4 \rangle  \right) \\
\end{eqnarray*}

Hence, the modular $\mathcal{S}$ matrix is given by

\begin{equation}
 \mathcal{S}  = \frac{1}{2} \left(
\begin{array}{cccc}
1 & 1 & 1 & 1 \nonumber \\
1 & 1 & -1 & -1 \nonumber \\
1 & -1 & 1 & -1 \nonumber \\
1 & -1 & -1 & 1
\end{array} \right)
\label{Smod} \end{equation}

This is exactly what one expects from the topological quantum field
theory corresponding to the zero correlation length
deconfined-confined $Z_2$ gauge theory. There are four flavors of
quasi-particles in the spectrum: $1, m, e, em$, as we have shown in
Table \ref{table1}. The electric charge $e$ and magnetic charge
(vison) $m$ both have self-statistics of a boson and pick up a phase
of $\pi$ when they encircle each other (and as a corollary, the same
phase when they encircle $em$). By studying $\mathcal{S}$, one gets
the self and mutual statistics for quasi-particles encircling each
other.

In Sec \ref{subsec:algo} we further show that symmetry is not
required to determine the $\mathcal{S}$ matrix. In Eqn.
\ref{toricmes} we have shown the MESs for cuts along $w_2 = \hat y$
direction:

\begin{eqnarray*}
|\Xi_1 \rangle = \frac{e^{i\varphi_1}}{\sqrt{2}}(|\xi_{00}\rangle + |\xi_{01}\rangle) \nonumber \\
|\Xi_2 \rangle = \frac{e^{i\varphi_2}}{\sqrt{2}}(|\xi_{00}\rangle - |\xi_{01}\rangle) \nonumber \\
|\Xi_3 \rangle = \frac{e^{i\varphi_3}}{\sqrt{2}}(|\xi_{10}\rangle + |\xi_{11}\rangle) \nonumber \\
|\Xi_4 \rangle = \frac{e^{i\varphi_4}}{\sqrt{2}}(|\xi_{10}\rangle -
|\xi_{11}\rangle)
\end{eqnarray*}

where $\varphi_i$ are undetermined phases for MESs $|\Xi_i \rangle$.
The unitary matrix $U_1$ connecting the $w_2$ MESs and the electric
flux states:

\begin{equation}
 U_1  = \frac{1}{\sqrt 2} \left(
\begin{array}{cccc}
e^{i\varphi_1} & e^{i\varphi_2} &  &   \\
e^{i\varphi_1} & -e^{i\varphi_2} &  &   \\
&  & e^{i\varphi_3} & e^{i\varphi_4}  \\
&  & e^{i\varphi_3} & -e^{i\varphi_4}
\end{array} \right)
\label{u1}
\end{equation}

On the other hand, it is straightforward to verify that for loops
along $w_{2}' = -\hat x + \hat y$ direction, which satisfies our
requirement Eqn.\ref{restrict2}, the corresponding MESs:

\begin{eqnarray*}
|\Xi_{1}' \rangle = \frac{e^{i\varphi_{1}'}}{\sqrt{2}}(|\xi_{00}\rangle + |\xi_{11}\rangle) \nonumber \\
|\Xi_{2}' \rangle = \frac{e^{i\varphi_{2}'}}{\sqrt{2}}(|\xi_{00}\rangle - |\xi_{11}\rangle) \nonumber \\
|\Xi_{3}' \rangle = \frac{e^{i\varphi_{3}'}}{\sqrt{2}}(|\xi_{01}\rangle + |\xi_{10}\rangle) \nonumber \\
|\Xi_{4}' \rangle =
\frac{e^{i\varphi_{4}'}}{\sqrt{2}}(|\xi_{01}\rangle -
|\xi_{10}\rangle)
\end{eqnarray*}

again $\varphi_{i}'$ are undetermined phases for MESs $|\Xi_{i}'
\rangle$. The unitary matrix $U_2$ connecting the $w_{2}'$ MESs and
the electric flux states:

\begin{equation}
 U_2  = \frac{1}{\sqrt 2} \left(
\begin{array}{cccc}
e^{i\varphi_{1}'}&  e^{i\varphi_{2}'}&  &  \\
&  &e^{i\varphi_{3}'}  & e^{i\varphi_{4}'} \\
&  &e^{i\varphi_{3}'}  & -e^{i\varphi_{4}'} \\
e^{i\varphi_{1}'}&  -e^{i\varphi_{2}'}&  &
\end{array} \right) \label{u2}
\end{equation}

Combining Eqn. \ref{u1} and \ref{u2}, we can write down the modular
$\mathcal S$ matrix as:

\begin{eqnarray*}
\mathcal S &=& U_{2}^{-1}U_{1} \nonumber \\ &=& \frac{1}{2} \left(
\begin{array}{cccc}
e^{i(\varphi_{1}-\varphi_{1}')}& e^{i(\varphi_{2}-\varphi_{1}')} & e^{i(\varphi_{3}-\varphi_{1}')} & -e^{i(\varphi_{4}-\varphi_{1}')} \\
e^{i(\varphi_{1}-\varphi_{2}')}& e^{i(\varphi_{2}-\varphi_{2}')} & -e^{i(\varphi_{3}-\varphi_{2}')} & e^{i(\varphi_{4}-\varphi_{2}')} \\
e^{i(\varphi_{1}-\varphi_{3}')}& -e^{i(\varphi_{2}-\varphi_{3}')} & e^{i(\varphi_{3}-\varphi_{3}')} & e^{i(\varphi_{4}-\varphi_{3}')} \\
e^{i(\varphi_{1}-\varphi_{4}')}& -e^{i(\varphi_{2}-\varphi_{4}')} &
-e^{i(\varphi_{3}-\varphi_{4}')}& -e^{i(\varphi_{4}-\varphi_{4}')}
\end{array} \right)
\end{eqnarray*}

To ensure the existence of an identity particle in accord with the
first row and column, we impose the conditions:

\begin{eqnarray*}
& & \varphi_{1}'= \varphi_{2}'= \varphi_{3}'= \varphi_{4}' \\
&=& \varphi_{1} = \varphi_{2} = \varphi_{3} = \varphi_{4}+\pi \\
\end{eqnarray*}

This leads to the following modular $\mathcal{S}$ matrix:

\begin{equation*}\mathcal{S}=\frac{1}{2}\left(\begin{array}{cccc}
1 & 1 & 1 & 1\\
1 & 1 & -1 & -1\\
1 & -1 & 1 & -1\\
1 & -1 & -1 & 1\end{array}\right)\end{equation*}

which is indeed the correct result for $Z_2$ toric code.

Now consider the transformation corresponding to $\mathcal{U}$ matrix as described in
Appendix \ref{sec:mod}, where $n'_1 = n_1$ and $n'_2 = n_2 - n_1$
are the winding numbers along the $\omega'_1$ and $\omega'_2$
directions. Using this expression and Eqn.\ref{toricmes}, the
transformation for MESs from $w_2$ cut to $w_2'$ cut:

\begin{eqnarray*}
|\Xi_1 \rangle &\rightarrow&  |\Xi_1 \rangle \\
|\Xi_2 \rangle &\rightarrow&  |\Xi_2 \rangle \\
|\Xi_3 \rangle &\rightarrow&  |\Xi_3 \rangle \\
|\Xi_4 \rangle &\rightarrow& -|\Xi_4 \rangle \\
\end{eqnarray*}

This leads to the following modular $\mathcal U$ matrix:

\begin{equation}
 \mathcal{U}  = \left(
\begin{array}{cccc}
1 & 0 & 0 & 0 \nonumber \\
0 & 1 & 0 & 0 \nonumber \\
0 & 0 & 1 & 0 \nonumber \\
0 & 0 & 0 & -1
\end{array} \right)
\label{Umod} \end{equation}

Again, this is what is expected from the $Z_2$ gauge theory. The
sign of $-1$ on the last entry of the diagonal corresponds to the
fermionic self statistics of the $em$ while the positive signs
correspond to the bosonic self statistics of $1, e$ and $m$
particles.

To see a more generic example to derive the $\mathcal{U}$ matrix
from rotation symmetry, we first define the toric code on a
triangular lattice, with system dimensions such that the $2 \pi/3$
rotation is a symmetry of the system. The Hamiltonian is same as Eq.
\ref{toriccode} with the star $`s'$ denoting six links emanating
from a vertex while the plaquette $`p'$ now involves three links. We
again denote the four degenerate ground states on a torus as
$|\xi_{ab}\rangle$ with $a,b = 0,1$ denoting the parity of electric
field along the non-contractible cycles. The relation between the
MESs $|\Xi_\alpha\rangle$ and the states $|\xi_{ab}\rangle$ remains
unchanged (Eqn. \ref{toricmes}). The calculation for the
transformation under $2 \pi/3$ proceeds analogously  to that for
$\pi/2$ rotation and one finds:

\begin{eqnarray*}
R_{2\pi/3} |\xi_{00}\rangle & = & |\xi_{00}\rangle \\
R_{2\pi/3} |\xi_{01}\rangle & = & |\xi_{10}\rangle \\
R_{2\pi/3} |\xi_{10}\rangle & = & |\xi_{11}\rangle \\
R_{2\pi/3} |\xi_{11}\rangle & = & |\xi_{01}\rangle \\
\end{eqnarray*}

Translating the action of $R_{2\pi/3}$ on the states
$|\xi_\alpha\rangle$ to that on states $|\Xi_\alpha\rangle$, one
finds

\begin{equation}
 \mathcal{US}  = \frac{1}{2} \left(
\begin{array}{cccc}
1 & 1 & 1 & 1 \nonumber \\
1 & 1 & -1 & -1 \nonumber \\
1 & -1 & 1 & -1 \nonumber \\
-1 & 1 & 1 & -1
\end{array} \right)
\label{USmod} \end{equation}

Combining the expression and the $\mathcal{S}$ matrix, one obtains

\begin{equation}
 \mathcal{U}  = \left(
\begin{array}{cccc}
1 & 0 & 0 & 0 \nonumber \\
0 & 1 & 0 & 0 \nonumber \\
0 & 0 & 1 & 0 \nonumber \\
0 & 0 & 0 & -1
\end{array} \right)
\label{Umod} \end{equation}

as expected.

\end{document}